\documentclass[review,12pt,times,authoryear]{elsarticle}

\usepackage{hyperref}
\hypersetup{
    colorlinks=false,
    linkcolor=black,
    filecolor=black,      
    urlcolor=black,
    pdftitle={PSU EFD},
    }

\usepackage{epsfig}
\usepackage{float}
\usepackage{subfig}
\usepackage{mathtools}
\usepackage{tabulary,xcolor}
\usepackage{booktabs}
\usepackage{lscape}
\usepackage{multirow}
\usepackage{amssymb}
\usepackage{changepage}
\usepackage{amsthm}
\usepackage{changepage}

\journal{}

\begin{document}

\begin{frontmatter}



\title{Surrogate Model for Shallow Water Equations Solvers with Deep Learning}


\author{Yalan Song}

\affiliation{organization={Department of Civil and Environmental Engineering, Pennsylvania State University},
            city={State College},
            postcode={16802}, 
            state={PA},
            country={United States}}
            
\author{Chaopeng Shen}

\affiliation{organization={Department of Civil and Environmental Engineering, Pennsylvania State University},
            city={State College},
            postcode={16802}, 
            state={PA},
            country={United States}}

\author{Xiaofeng Liu}

\affiliation{organization={Department of Civil and Environmental Engineering, Institute of Computational and Data Science, Pennsylvania State University},
            city={State College},
            postcode={16802}, 
            state={PA},
            country={United States}}

\begin{abstract}
Shallow water equations are the foundation of most models for flooding and river hydraulics analysis. These physics-based models are usually expensive and slow to run, thus not suitable for real-time prediction or parameter inversion. An attractive alternative is surrogate model. This work introduces an efficient, accurate, and flexible surrogate model, NN-p2p, based on deep learning and it can make point-to-point predictions on unstructured or irregular meshes. The new method was evaluated and compared against existing methods based on convolutional neural network (CNNs), which can only make image-to-image predictions on structured or regular meshes. In NN-p2p, the input includes both spatial coordinates and boundary features that can describe the geometry of hydraulic structures, such as bridge piers. All surrogate models perform well in predicting flow around different types of piers in the training domain. However, only NN-p2p works well when spatial extrapolation is performed. The limitations of CNN-based methods are rooted in their raster-image nature which cannot capture boundary geometry and flow features exactly, which are of paramount importance to fluid dynamics. NN-p2p also has good performance in predicting flow around piers unseen by the neural network. The NN-p2p model also respects conservation laws more strictly. The application of the proposed surrogate model was demonstrated by calculating the drag coefficient $C_D$ for piers and a new linear relationship between $C_D$ and the logarithmic transformation of pier's length/width ratio was discovered.
\end{abstract}


\begin{highlights}
\item A surrogate model for shallow water equations solver was developed using a point-to-point prediction approach.
\item The new model overcomes the limitations in raster-image based approaches. 
\item The model can successfully predict flow fields and respect physical laws with high accuracy.
\end{highlights}


\end{frontmatter}



\section{Introduction}
With climate change, floods have become one of the most common natural hazards which affect hundreds of millions of people worldwide and cost billions of dollars of damage annually \citep{NWS2020,Cigler2017}. Accurate and real-time flood forecasting and predictions are of great importance to increase resilience and reduce damage. The backbone of most flood simulation models is the two-dimensional (2D) shallow water equations (SWEs), also known as the Saint-Venant equations \citep{Liu2008}. They are a class of hyperbolic partial differential equations (PDEs) which approximate the full three-dimensional motion of fluid as two-dimensional. Many 2D SWEs solver have been developed, which typically use traditional methods such as the finite volume method (FVM) to discretize the governing equations on a mesh. Such solvers are often called physics-based models (PBMs). Despite their wide use and prevalence in practice, PBMs are usually computationally expensive. Thus, they are not suitable for real-time predictions or parameter inversion which needs large number of runs.



One attractive alternative to PBMs is the computationally efficient surrogate models, which approximate the input-output dynamics of the system. With the recent renaissance of machine learning (ML) and artificial intelligence (AI), surrogate modeling approach has gained substantial popularity. There are many ways to construct surrogate models based on data and the underlying physics. Deep learning (DL) \citep{lecun2015deep,shen2018transdisciplinary}, especially deep neural network (DNN), is one of the most popular ways for building surrogate models due to its potential for capturing high-dimensional nonlinearity. There are also other surrogate model building techniques, such as the stochastic approach using Gaussian processes \citep{bilionis2012multi,bilionis2013multi,zhu2018bayesian,coppede2019hydrodynamic}, 
the reduced-order models or intermediate-complexity models which have  less complexity than high-fidelity models \citep{lang2009reduced,amsallem2015design}, and the hybrid physics-ML models to combine physics and data-driven methods \citep{tracey2015machine,wu2018physics,duraisamy2021perspectives,zhu2019machine,kurz2020machine,duraisamy2021perspectives}. This work uses DNN to build surrogate models which can accurately and efficiently capture show flow dynamics. In addition, surrogate models built with DNN supports automatic differentiation, i.e., the automatic calculation of the gradient of output with respect to input. Such gradient information is extremely useful for solving previously difficult or unsolvable problems. For example, \cite{tsai2021calibration} recently proposed a novel differentiable parameter learning (dPL) framework to integrate big-data DL and differentiable PBMs for parameter calibration. Another example is the use deep learning surrogate model to perform riverine bathymetry inversion \citep{GHORBANIDEHNO2021}.

To appreciate the difference between traditional solvers and ML/AI based approaches for SWEs, it is beneficial to have a brief survey of their respective landscapes. Over the past several decades, 2D hydraulic models are increasingly used in academic researches and engineering practice, such as SRH-2D by U.S. Bureau of Reclamation (USBR) \citep{lai2010two}, HEC-RAS 2D by U.S. Army Corp of Engineers (USACE) \citep{brunner1995hec}, FLO-2D \citep{o2011flo}, RiverFlow2D \citep{hydronia2016riverflow2d}, MIKE 21 \citep{warren1992mike}, TUFLOW \citep{huxley2016tuflow}, LISFLOOD \citep{van2010lisflood}, among many others. The general trend in this realm is solvers are getting more complicated by incorporating more physics and increasingly available high resolution terrain and hydrometeorological data. They are also getting faster with the use of parallel computing techniques such as OpenMP, Message Passing Interface (MPI), and General-Purpose Graphics Processing Unit (GPGPU). Despite the advancements, they are still slow for applications when time is of essence or large number of runs are needed. 

On the other hand, ML/AI techniques have been used to solve flow governing equations, such as the Navier-Stokes equations and SWEs \citep{tompson2017accelerating,raissi2018hidden,qian2019physics,lapeyre2020reconstruction,forghani2021application,bihlo2021physics}. These techniques can be classified into two categories: unsupervised methods and supervised methods. Physics-informed neural network (PINN) is a popular unsupervised method, in which governing equations, boundary conditions (BCs), and initial conditions (ICs) for flow are enforced in the loss function \citep{raissi2019physics,rao2020physics,mao2020physics,bihlo2021physics}. The training data used to minimize the loss function only consist of the coordinates of sample points from the internal field and boundaries. The outputs of PINN models are made of flow variables such as velocity, pressure and water depth, which are plugged into governing equations and BCs/ICs to calculate error (loss). The supervised methods are also called data-driven methods or surrogate models. A large set of representative flow data, mostly from PBM simulations, is needed to train the model. The physics and constrains, i.e., governing equations and BCs/ICs, are implicitly embedded in the data and to be learned by ML algorithms.

This work uses supervised learning to build a surrogate emulating 2D SWEs solvers. The solutions of 2D hydraulics models are the distribution of flow variables on a 2D plane, typically in the form of images such as inundation maps. One important strength of ML/AI is image processing and recognition. In the context of this work, previous researches have used image-to-image regression to build surrogate models \citep{zhu2018bayesian}, which map domain boundary images to flow solution images. Among many ML techniques, convolutional neural network (CNN) is a very popular method for image-to-image regressions \citep{guo2016convolutional,bhatnagar2019prediction}. For example, the input of CNN provides the boundary condition with a image, which can be viewed as a matrix of a binary function (boundary pixel or external) or a signed distance function (SDF) on the Cartesian grid. \citep{guo2016convolutional}. The outputs are flow variables on the same Cartesian grid (image pixels). High-accuracy flow predictions have been reported using CNN. 

CNN-based models can only be used on structured or regular (Cartesian) mesh. The prediction is constrained in the domain where the image pixels (the Cartesian grid) are defined. Thus, CNN-based surrogates theoretically cannot predict flow outside the training domain. In addition, their training time increase exponentially with the increase of image resolution. However, high-resolution images, i.e., high-resolution results, are always desired in practice. To exacerbate the problem, it is well known that raster images are not well suited to represent linear features such as bridge piers, roads, levies, and other boundaries which steer the flow. Perhaps more importantly, features such as sharp corners, which controls flow separation and overall hydrodynamics, need extremely high-resolution raster images to be captured. 

Improvements to overcome the inherent weakness of raster images have been proposed. Mostly, the main idea is to introduce vectorized information in the training process. For example, \cite{sekar2019fast} proposed to combine CNN and Multilayer Perceptron (MLP) network for the prediction of flow around airfoils. They used CNN to preprocess the boundary information of airfoil into several geometric features. Then, the MLP network was used for a point-by-point prediction of flow fields. The point coordinates, geometric features of the airfoil, and flow conditions are the inputs of the MLP network. Flow variables at input point are the outputs. This method has better performance in the near object region because the points on airfoil boundaries, rather than the nearest points to the boundaries on the Cartesian grid, are used in the training. Despite its success, the combined use of CNN and MLP increases the complexity and training cost of ML model.

In this work, shallow open channel flow around structures, such as bridge piers, is studied, which is critical for infrastructure safety and resilience against floods. Different from \cite{sekar2019fast} which uses CNN to indirectly embed object boundaries in the training, we directly specify the geometric features of bridge piers as part of the input. Geometric features of piers include their shape and size.  From the practical point of view, the efficacy of out surrogate model is evaluated by answering the following questions:
\begin{itemize}
  \item How is the prediction of the surrogate model for flow around training piers but with unseen dimensions (length and width)?
  \item How is the prediction of the surrogate model for flow around piers with unseen shapes? This is inherently more difficult than the previous question.
  \item Can the surrogate model accurately predict the flow outside the training zone? 
  \item How well does the surrogate model respect physical laws such as the conservation of mass?
\end{itemize}

As will be shown in this paper, our new method performs better than or at least equally well as previous methods. We will demonstrate the new method's performance by answering the above questions. We will further demonstrate the application of the surrogate model by evaluating the accuracy of the produced drag coefficient, a key parameter for the safety and longevity of bridges. 


\section{Methodology}
This section first describes the shallow water equations and the PBM solver used in this work. Then, the surrogate model using deep learning is introduced. For comparison purpose, two CNN-based surrogate models are also briefly described.

\subsection{Shallow water equations and solver}
The shallow water equations can be derived by depth-averaging the 3D Navier-Stokes equations. They have the following general form \citep{lai2010two}:
\begin{equation}\label{eqn:cty}
\frac{\partial h}{\partial t} + \frac{\partial h u}{\partial x} + \frac{\partial h v}{\partial y} =0
\end{equation}
\begin{equation}\label{eqn:momentum_x}
\frac{\partial h u}{\partial t} + \frac{\partial h u u}{\partial x} + \frac{\partial h u v}{\partial y} =  \frac{\partial h {T_{xx}}}{\partial x}+\frac{\partial h {T_{xy}}}{\partial y}-gh\frac{\partial z_s}{\partial x}-\frac{\tau_{bx}}{\rho}
\end{equation}
\begin{equation}\label{eqn:momentum_y}
\frac{\partial h v}{\partial t} + \frac{\partial h uv}{\partial x} + \frac{\partial h v v}{\partial y} =  \frac{\partial h{T_{xy}}}{\partial x}+\frac{\partial h {T_{yy}}}{\partial y}-gh\frac{\partial z_s}{\partial y}-\frac{\tau_{by}}{\rho}
\end{equation}
where $u$ and $v$ are the depth-averaged flow velocities in $x$ and $y$ directions, respectively; ${h} $ is the water depth; $z_s = h+z_b$ is the water surface elevation; $z_b$ is the bed elevation;  ${T_{xx}}$,${T_{xy}}$, and ${T_{xy}}$ are the depth-averaged turbulence stresses; $\tau_{bx}$ and $\tau_{by}$ are the bed shear stresses in $x$ and $y$ directions, respectively; $g$ is the gravitational acceleration; $\rho$ is the water density. More details about the turbulence model and the physical means of all terms can be found in \cite{rodi1993turbulence} and \cite{lai2008srh}.

To generate the training data, the SRH-2D model was used to simulate flow fields around bridge piers. SRH-2D is a popular 2D hydraulics model which solves the SWEs using FVM on unstructured meshes \citep{lai2008srh}. Because of the large number of simulations needed, and to automate the data generation and processing, the python package, Python-based Hydraulic Modeling Tools, $pyHMT2D$, was used to control SRH-2D modeling runs and transform the results to the inputs and outputs of the neural network \citep{pyHMT2D}. Source code of $pyHMT2D$ can be found on GitHub (\href{https://github.com/psu-efd/pyHMT2D}{https://github.com/psu-efd/pyHMT2D}). 

Drag force $F_D$ on bridge piers can be calculated from the simulated flow field. Indeed, the drag force is dominated by the pressure force acting on the surface of pier and a drag coefficient $C_D$ can be calculated as:
\begin{equation}
C_D=\frac{F_D}{0.5 A_d \rho U^2}=\frac{\int_\mathbf{A_w} p d \mathbf{A_w}}{0.5 A_d \rho U^2}
\end{equation}
where $\rho$ is the density of water, $U$ is the average velocity of incoming flow, $A_d$ is the projected area of pier on a plane perpendicular to the flow. $A_w$ is the wetted area of pier. Because of the hydrostatic assumption in SWEs, the pressure $p$ at any point on a bridge pier is a linear function of its water depth.

\subsection{Surrogate models}
\begin{figure}[H]
\centering
    \includegraphics[width=0.8\textwidth]{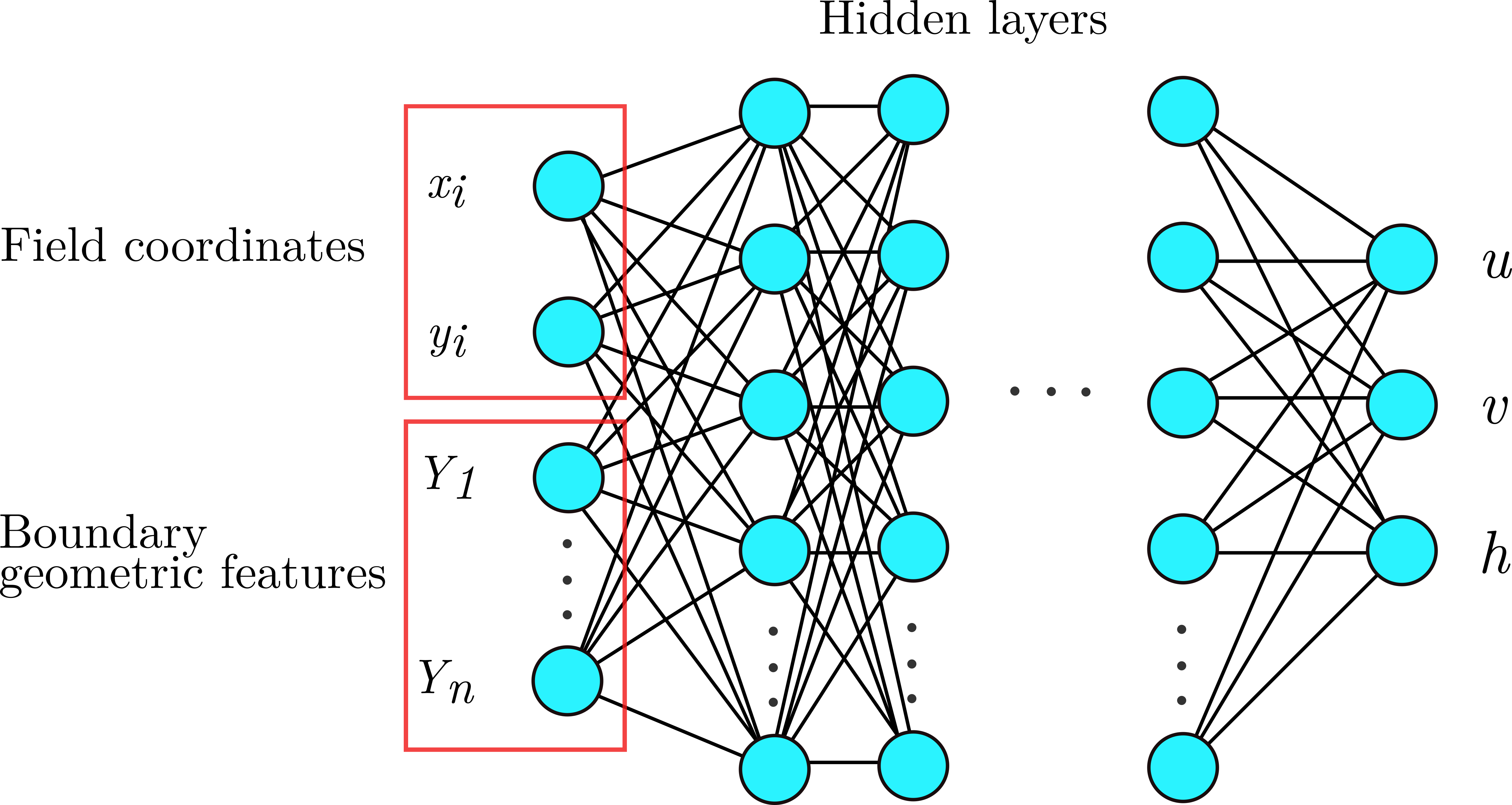}
    \caption{Scheme diagram of a multi-layer perceptron neural network for the surrogate model.}\label{Fig:fully_connected_layer}
\end{figure}

We proposed a MLP neural network to represent the nonlinear solutions of SWEs as shown in Fig.~\ref{Fig:fully_connected_layer}. The flow field governed by SWEs is a complex nonlinear function of coordinates ($x_i$ and $y_i$), boundary geometric features of the piers (details next), discharge and water depth conditions at upstream and downstream boundaries, etc. In this work, we limit our focus on the effects of pier shape and size to the flow at steady state. Other parameters, such as discharge, water depth, Manning's n, and channel shape, are the same for all data, which can be considered as additional input parameters in future study. Thus, the input explanatory variables in this work only include a pair of grid coordinates  ($x_i$ and $y_i$) and boundary geometric features. The output response variables are the velocities ($u$ and $v$) and water depth, $h$, at the input point. This method is based on point-to-point regression using neural network, noted as NN-p2p in this work. 

For comparison, image-to-image regression methods based on CNN were also implemented. In these methods, the input images are used to represent the pier boundaries in the domain. Two different methods to generate the input images were tested. In one method, the boundaries are represented with a binary field (1 on the boundary and 0 otherwise). In the second method, the boundaries are represented with a signed distance function (SDF) where each pixel's value is its distance to the nearest boundary. To distinguish the two, they are noted as CNN-binary and CNN-SDF, respectively. More details on how to generate the matrices (input images) can be found in \cite{guo2016convolutional}. 

\begin{figure}[H]
\hspace{-60pt}
    \includegraphics[width=1.2\textwidth]{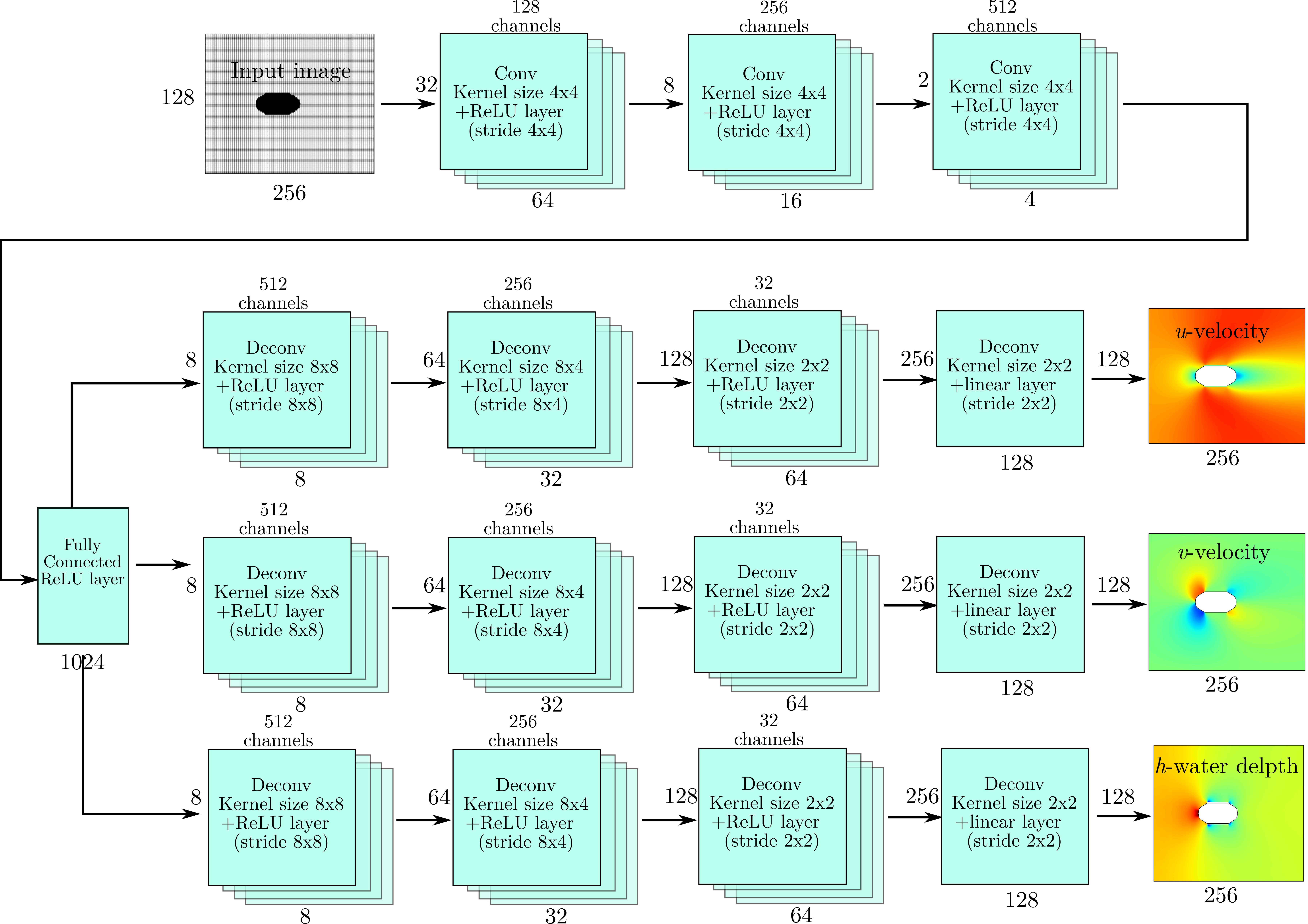}
    \caption{Scheme diagram of the CNN for image-to-image regression.}\label{Fig:CNN_structure}
\end{figure}

In the two CNN-based methods, the input images are encoded to a geometric feature vector by multiple convolution layers and a fully connected layer. The geometric feature vector contains the information that represents the input images, i.e., the boundaries. Then, the decoder maps the geometric feature vector to three output images through multiple deconvolution layers. The three output images are for $\hat{u}$, $\hat{v}$ and $\hat{h}$, respectively. These output flow images are then compared with the simulated images (ground truth) from SRH-2D to calculate loss. The scheme diagram of the CNN-based methods is shown in Fig.~\ref{Fig:CNN_structure}.

The loss function in all surrogate models is defined as the mean square error (MSE) of prediction:
\begin{equation}
E_{loss}=\frac{1}{3} \frac{1}{N_{batch}}  \frac{1}{M_{sample}} \sum_{n=1}^{N_{batch}}\sum_{m=1}^{M_{sample}} \left[ (\hat{{u}}-{u})^2+ (\hat{{v}}-\overline{v})^2 + (\hat{h}-h)^2 \right]
\end{equation} 
where $u$, $v$, and $h$ are the flow data (ground truth) from SRH-2D simulations. $\hat{u}$, $\hat{v}$, and $\hat{h}$ are the predicted flow data from the surrogate models. $N_{batch}$ is the number of batches and $M_{sample}$ is sample size in each batch. $n$ and $m$ are the index for batch and sample, respectively. A stochastic gradient descent optimizer, ADAM, was used to update the weights and biases of the neural networks \citep{kingma2014adam}. The ReLU activation function was utilized to alleviate the vanishing gradient problem, which showed much better performance than the the hyperbolic tangent (\textit{tanh}) activation function used in \cite{sekar2019fast}. For the output layer, the linear activation function is used to accommodate negative output values. 

The performance of surrogate models highly depends on the choice of hyperparameters. The optimal values for these hyperparameters were obtained through manual tuning. For the NN-based surrogate model proposed in this work, it contained 10 hidden layers and each layer contains 1000 neurons. For the CNN-based surrogate models, the CNNs consisted of a 3-layer encoder and a 4-layer decoder. The kernel size and other hyperparameters are shown in Fig.~\ref{Fig:CNN_structure}. To update the weights and biases in the neural networks, an initial learning rate was set at $10^{-4}$. A learning rate scheduler was utilized to reduce the learning rate by a factor of 0.5 when the MSE of the validation dataset did not decrease over 3 epochs.

The surrogate models and their neural networks were implemented in Tensorflow \citep{tensorflow2015-whitepaper} with Keras API \citep{chollet2015keras}. The training of the surrogate models was performed on Amazon AWS with a NVIDIA K80 GPU card. The simulations with the physics-based model SRH-2D were performed on a desktop with an Intel Core 3.40GHz CPU. The code and data used in this work can be accessed at \href{https://github.com/psu-efd/Surrogate_Modeling_SWEs}{https://github.com/psu-efd/Surrogate\_Modeling\_SWEs}.

%

\section{Data Generation and Preprocessing}
\begin{figure}[H]
\centering
     \subfloat[][]{
    	\includegraphics[width=0.4\textwidth]{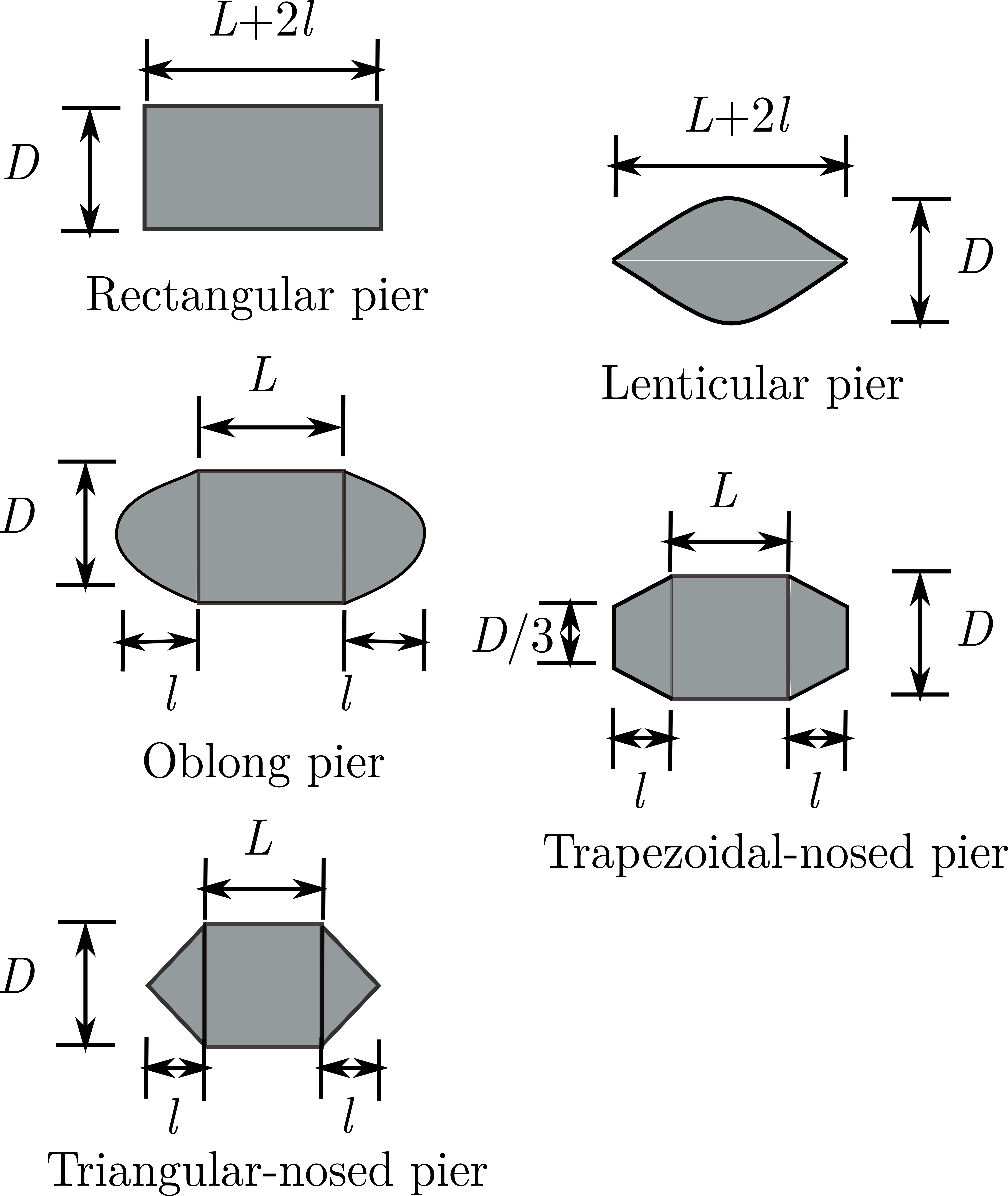}
     }      
     \subfloat[][]{
    	\includegraphics[width=0.4\textwidth]{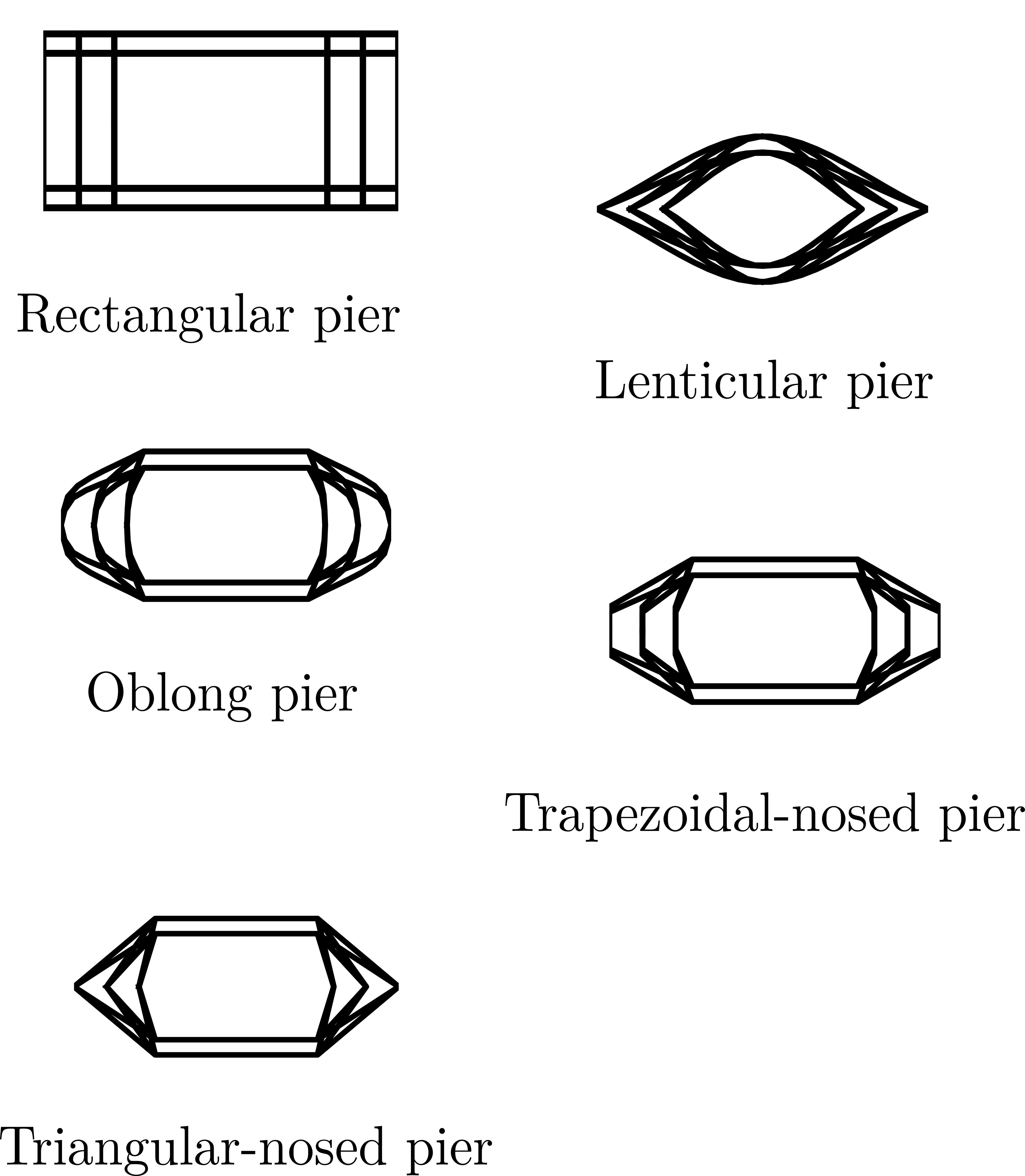}
     }      
    \caption{Piers uses in this work: (a) five different shapes, including rectangular, oblong, triangular-nosed, lenticular and  trapezoidal-nosed; (b) show the boundaries of piers of different shapes when $l$ is 0.1 m, 0.3 m, and 0.5 m, and $D$ is 0.1 m and 0.2 m.}\label{Fig:Piers_boundaries}
\end{figure}

For surrogate models to learn the effects of pier on flow, our training data include simulated flow around five different types of piers: rectangular, oblong, triangular-nosed, lenticular and trapezoidal-nosed shapes (see Fig.~\ref{Fig:Piers_boundaries} (a)). The total length of pier, $L_p=L+l$, changes from 1.04 m to 2.08 m, in which $L$ has a constant value of 1 m and $l$ ranges from 0.02 m to 0.54 m with a step size of 0.02 m. The width of the pier, $D$, ranges from 0.54 m to 1.06 m with a step size of 0.04 m. For each shape, Figure~\ref{Fig:Piers_boundaries} (b) shows some example boundaries of piers when $l$ is in the set [0.1 m, 0.3 m, 0.5 m] and $D$ is in the set [0.1 m, 0.2 m]. Considering the combinations, a total of 1,890 SRH-2D simulations were performed for five pier shapes. The piers were positioned at the center of the channel, which is 15 m long and 6 m wide. 

\begin{figure}[htp]
\centering
    \includegraphics[width=0.8\textwidth]{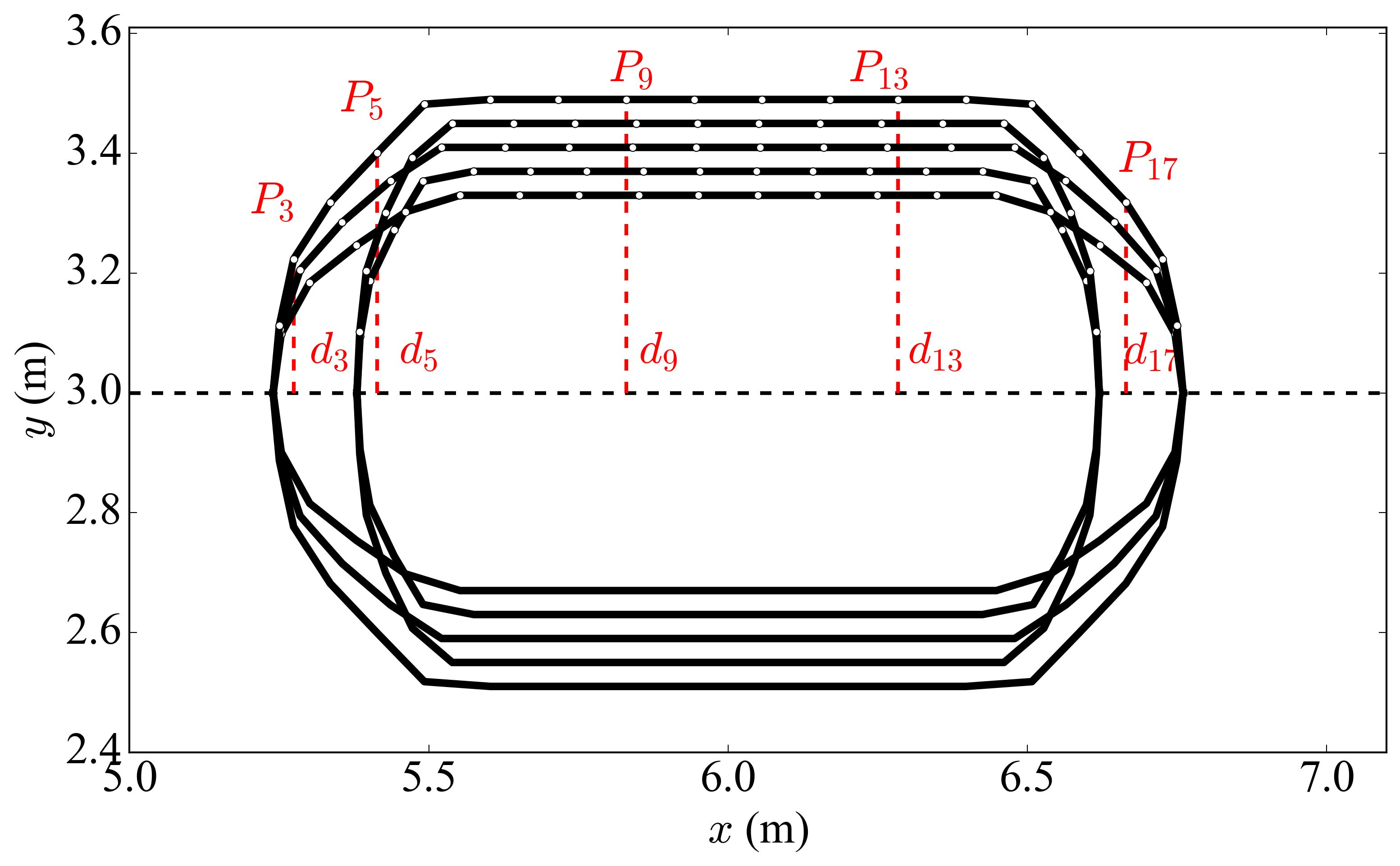}
    \caption{Boundary geometric features of piers. Points are sampled on the boundary of each pier to sketch the outline. In this work, eighteen points are used for each pier. Not all points are shown.}\label{Fig:boundary_features}
\end{figure}

In the NN-p2p model, it is important to design unique features as part of inputs to distinguish piers of different shapes and dimensions. Geometrically, these five piers have different shapes and it is hard to distinguish them using a common set of parameters such as length, width, curvature, etc. For example, the oblong shape has two round caps which need a radius of curvature. In contrast, the rectangular shape do not need that. To overcome this problem, we propose an approximate method for constructing pier's boundary geometric features ($Y_1$, $Y_2$, ...,  $Y_n$ in Fig.~\ref{Fig:fully_connected_layer}). In this method, a fixed number of points ($P_1$, $P_2$, ...,  $P_n$) along the boundary of the pier are used to describe the pier as shown in Fig.~\ref{Fig:boundary_features}. The curve length between adjacent points is uniform to capture the curvature of the boundary. As a result, the boundary geometric feature is a vector ($Y_1$, $Y_2$, ...,  $Y_n$), whose elements are the distances ($d_1$, $d_2$, ...,  $d_n$) of points ($P_1$, $P_2$, ...,  $P_n$) to the symmetry plane normalized by half width $D/2$. As in any approximation method, the more points, i.e., the larger $n$ is, the better the piers will be distinguished. In this work, we found the use of 18 points was sufficient for the cases studied in this work.

\begin{figure}[htp]
\centering
	 \subfloat[][]{
    	\includegraphics[width=0.8\textwidth]{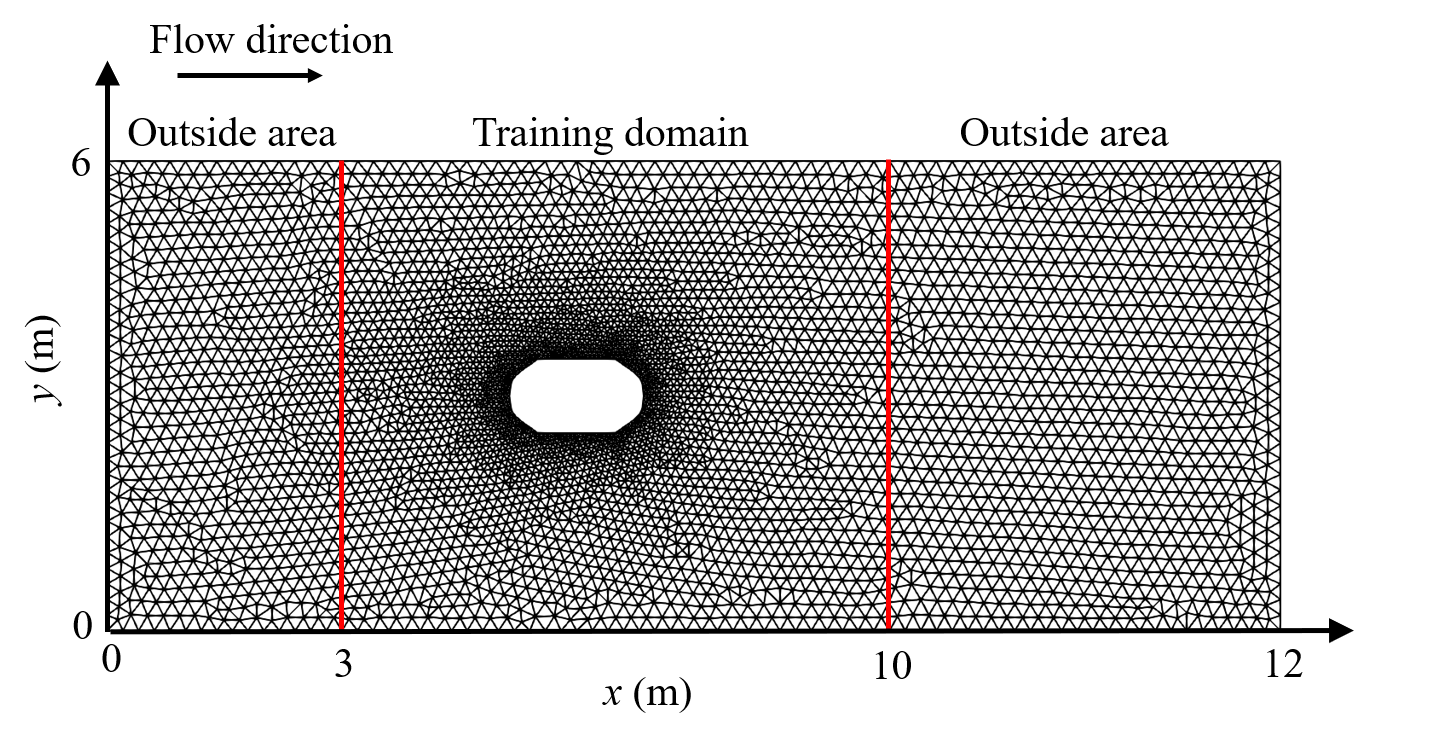}
     }    
       
     \subfloat[][]{
    	\includegraphics[width=0.45\textwidth]{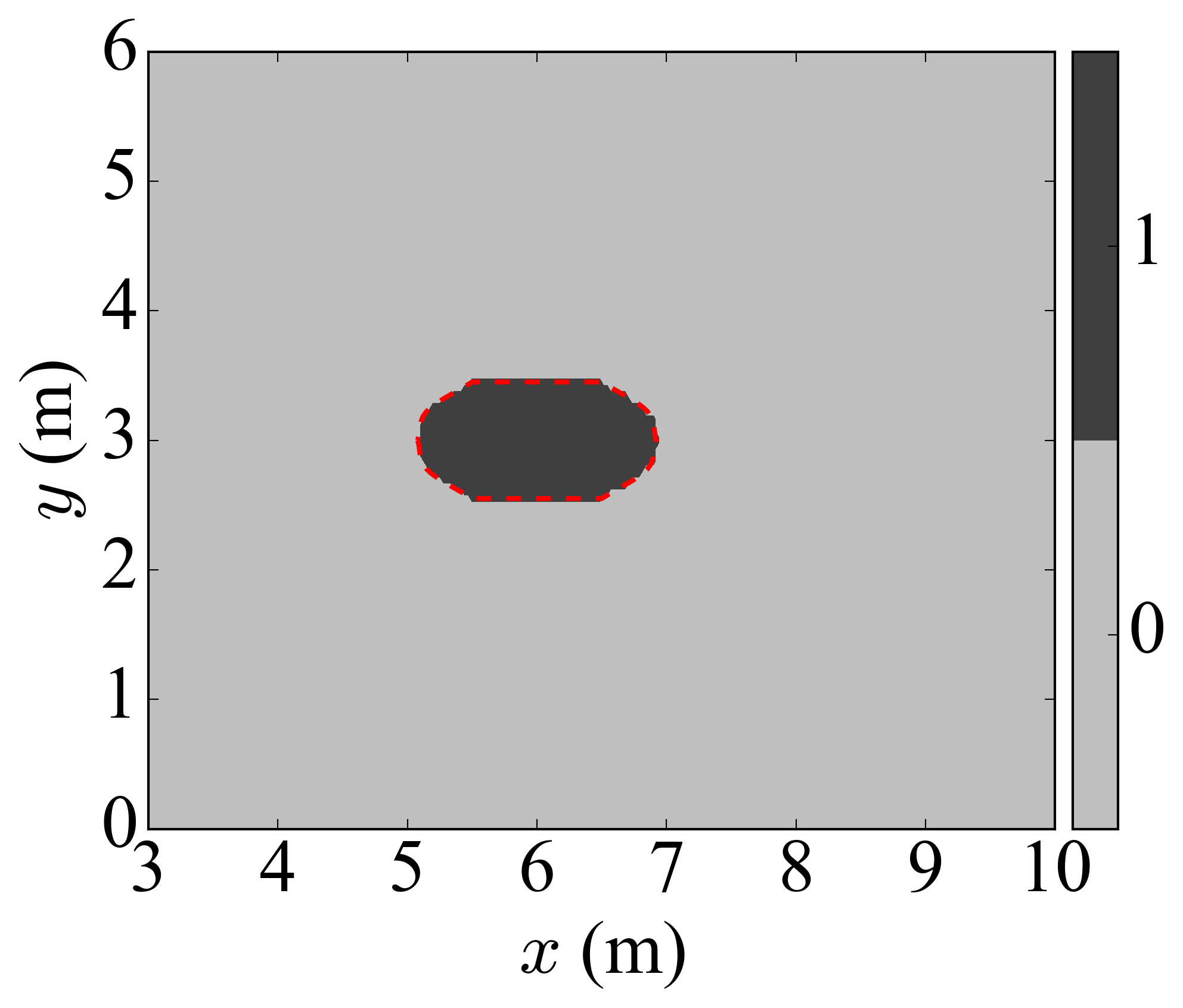}
     } 
	\subfloat[][]{
    	\includegraphics[width=0.46\textwidth]{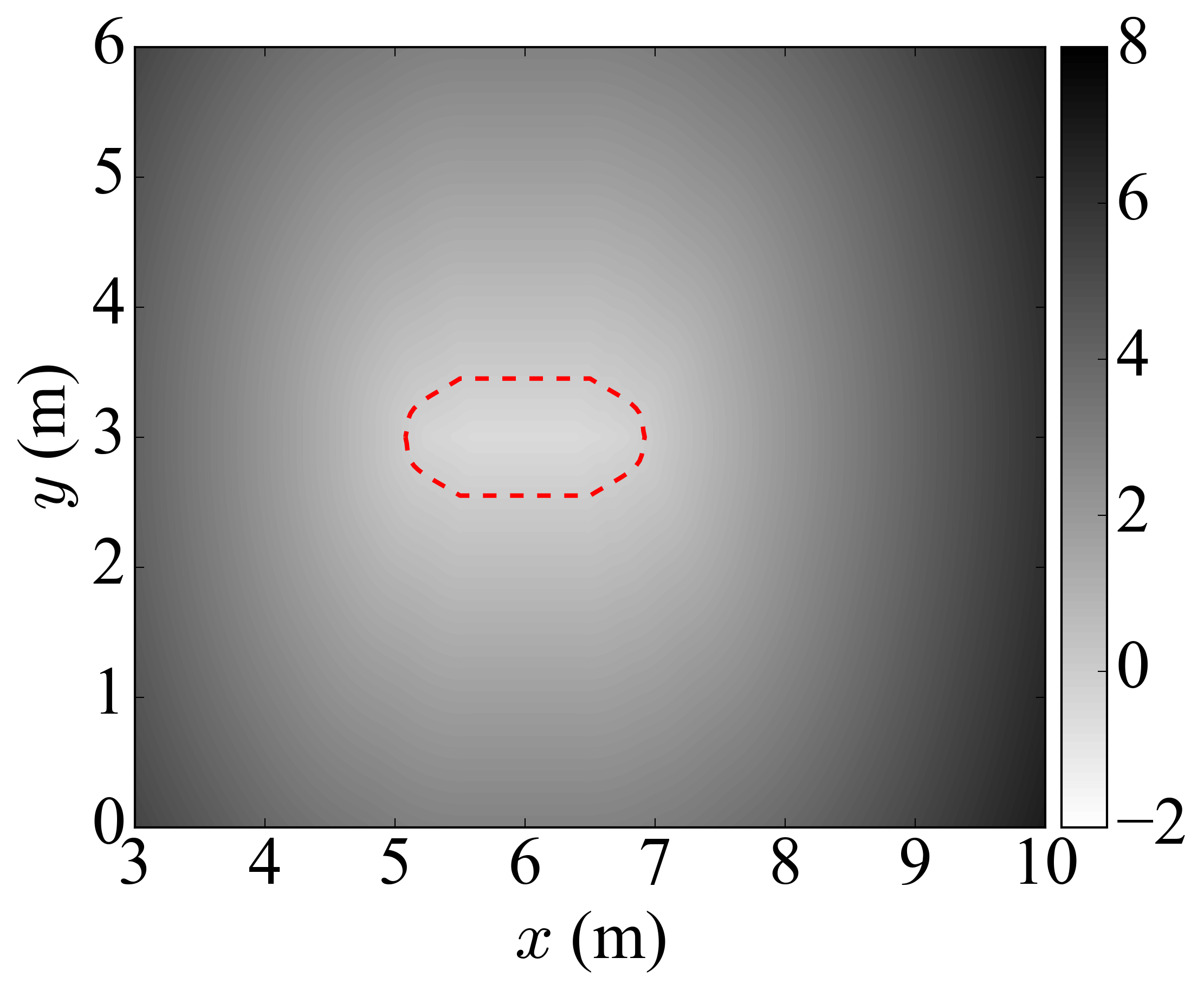}
     }    
    \caption{Simulation domain: (a) schematic view of an example mesh used for flow around an oblong pier. The red lines are the boundaries for training domain. (b) An example input image for CNN-binary. (c) An example input image for CNN-SDF where the red dashed line is for the pier.}\label{Fig:mesh}
\end{figure}

Fig.~\ref{Fig:mesh} shows a schematic view of the mesh used in an SRH-2D simulation (an oblong pier is used as an example). The coordinate origin is set at the lower-left corner of the domain. The mesh resolution is 0.2 m and 0.05 m in the far field and the near field (around object), respectively. In all simulations, the incoming flow specified at the left inlet boundary is 12 m$^3$/s and the water depth at the right outlet boundary is set as 2 m. As a result, the average velocity $U_0$ in the domain is about 1 m/s and the Reynolds number, $Re_D = U_0 D/\nu$ is 5 $\times$ 10$^5 \sim$ 10$^6$. To reduce the training cost, not all simulation data were used. Instead, the data were extracted from the training domain around the pier as shown in Fig.~\ref{Fig:mesh} (a). The training domain is defined as 3 m $<x<$ 10 m and 0 m $<y<$ 6 m. We randomly selected 60\% simulations for each pier shape as training data, 20\% simulations as cross-validation data, and 20\% simulations as testing data. Each simulation contains about 3,300 data points for NN-p2p. The total dataset of NN-p2p consists of about 6.1 million samples. The batch size of NN-p2p is 2048. 

In CNN models, each simulation only contains one sample with an input image and three output images. The total dataset of CNNs consists of 1,890 samples. The batch sizes of CNN models are 20. The input image of CNN-binary (Fig.~\ref{Fig:mesh} (b)) is an matrix of a binary field. When a pixel is inside the pier (red dashed line), its value is 1. Otherwise, it is 0. The input image of CNN-SDF (Fig.~\ref{Fig:mesh} (c)) is a matrix of the signed distance function. When a pixel is inside the pier, it is negative and its magnitude is the distance of the pixel to the nearest boundary. Otherwise, the distance value is positive. The pixel resolution of all images is 256 in $x$ direction and 128 in $y$ direction. The resolution of the input image is high enough to capture the curvature of the pier boundary.

\section{Results and discussions}

\subsection{Evaluation and comparison of surrogate models}
\begin{figure}[H]
\hspace{-50pt}
    	\includegraphics[width=1.2\textwidth]{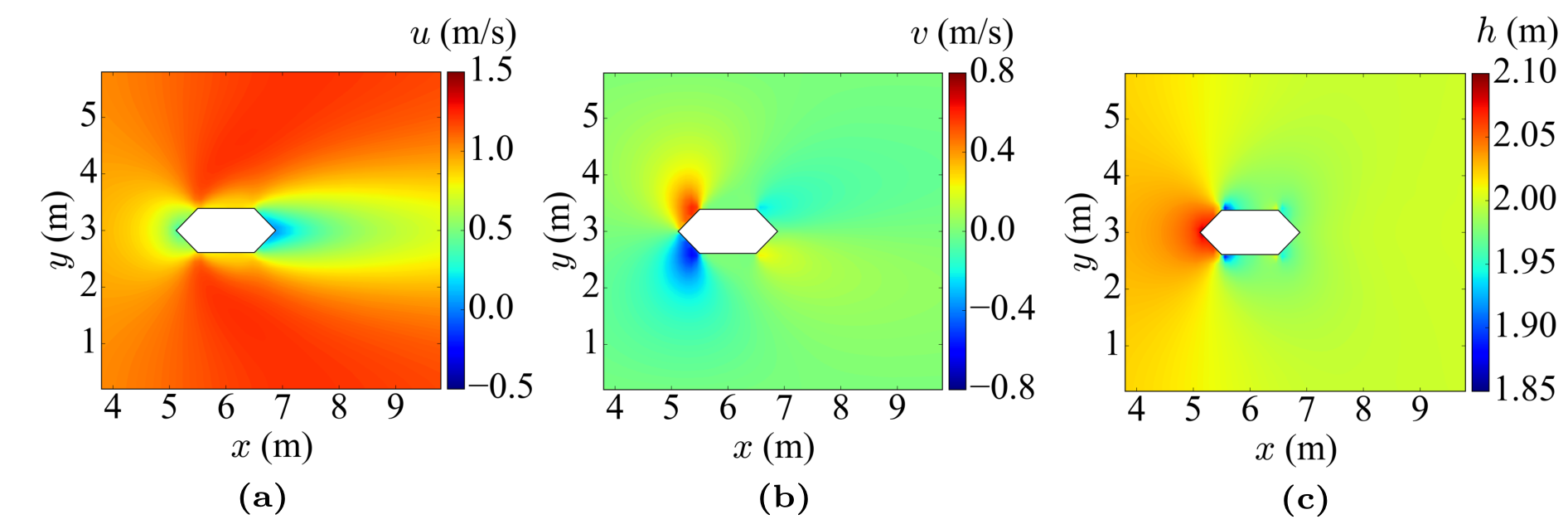} 
    \caption{Flow and water depth contours around a triangular-nosed pier simulated by SRH-2D: (a) streamwise velocity $u$, (b) transverse velocity $v$, (c) water depth $h$.}\label{Fig:tri_nosed_pier_simulation}
\end{figure}

In this section, the three surrogate models, NN-p2p, CNN-binary, and CNN-SDF, are evaluated and compared. It is impossible to show every test cases in the dataset. Instead, a case for triangular-nosed pier with unseen dimensions ($D$ = 0.78 m, $L_p$ = 1.76 m) from the testing data is demonstrated as an example. All other cases show similar model behavior. To have an overall view of the flow field, Fig.~\ref{Fig:tri_nosed_pier_simulation} shows the contours of $u$, $v$ and $h$ from SRH-2D simulation. It is clear that the three solution variables have very different distribution and range around the pier. The streamwise velocity $u$ dominates over the transverse velocity $v$ in the domain, except in the area near the front nose where flow is steered to the side. The relative change of water depth is small in comparison with the change of velocity. Because of all these, the error in $u$ is dominant in the total loss function. Consequently, the trained model predicts $u$ better than $v$ and $h$. 

The training data were normalized to reduce the dominance of $u$ and boost the contributions from $v$ and $h$. In addition, the spatial variations of solution variables do not follow the same pattern. In particular, Fig.~\ref{Fig:tri_nosed_pier_simulation} shows that the variations in $v$ and $h$ are mainly in the region close to the pier and vanish quickly as the distance to the pier increases. On the other hand, the variation of $u$ is more widespread. Our experience in performing ML training show that such cases are prone to vanishing gradient problem. One symptom of such problem is the predicted flow field near the pier does not show the fast variations. The use of the ReLU activation function greatly alleviated this problem. 



The surrogate models were evaluated with two metrics, i.e., the average absolute and relative errors. The relative error for a grid point is defined as the ratio of the absolute error to the maximum of the predicted and ground-truth values (to avoid divided by zero problem). Taking $u$ as an example, the average absolute and relative errors ${Ea}_{{u}}$ and ${Er}_{{u}}$ are defined as follows:
\begin{equation}
\begin{aligned}
{Ea}_{{u}} &= \frac{1}{N_{cases}}  \sum_{i=1}^{N_{cases}} \left( \frac{1}{N_{points}}  \sum_{i=1}^{N_{points}}  \hat{{u}}-{u} \right)\\ 
{Er}_{{u} }&= \frac{1}{N_{cases}}  \sum_{i=1}^{N_{cases}} \left( \frac{1}{N_{points}}  \sum_{i=1}^{N_{points}}  \frac{\hat{{u}}-{u}}{\max(\hat{u},u)} \right)\\ 
\end{aligned}\label{eqn:errors}
\end{equation}
where $N_{cases}$ is the number of cases in the testing dataset and $N_{points}$ is the number of points in each case. Because the variation of water depth is very small, we calculated the errors for water depth change $\Delta h$, rather than $h$. 

\begin{figure}[H]
\hspace{-50pt}
    	\includegraphics[width=1.15\textwidth]{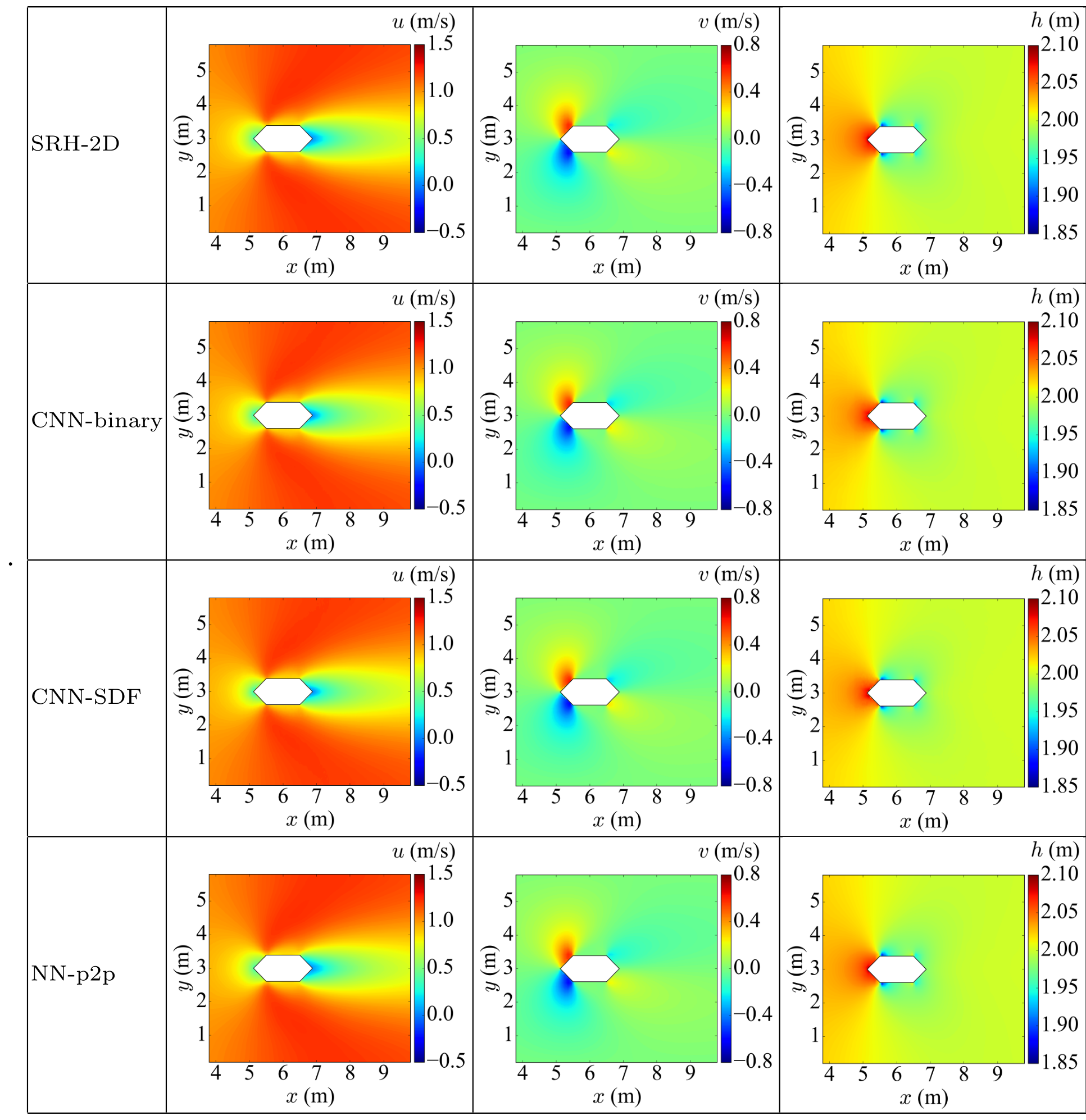} 
  \caption{Velocity and water depth contours around a triangular-nosed pier from SRH-2D simulation, CNN-binary, CNN-SDF and NN-p2p.}\label{Fig:tri_nosed_pier_prediction}
\end{figure}
\begin{figure}[htp]
\centering
    	\includegraphics[width=0.7\textwidth]{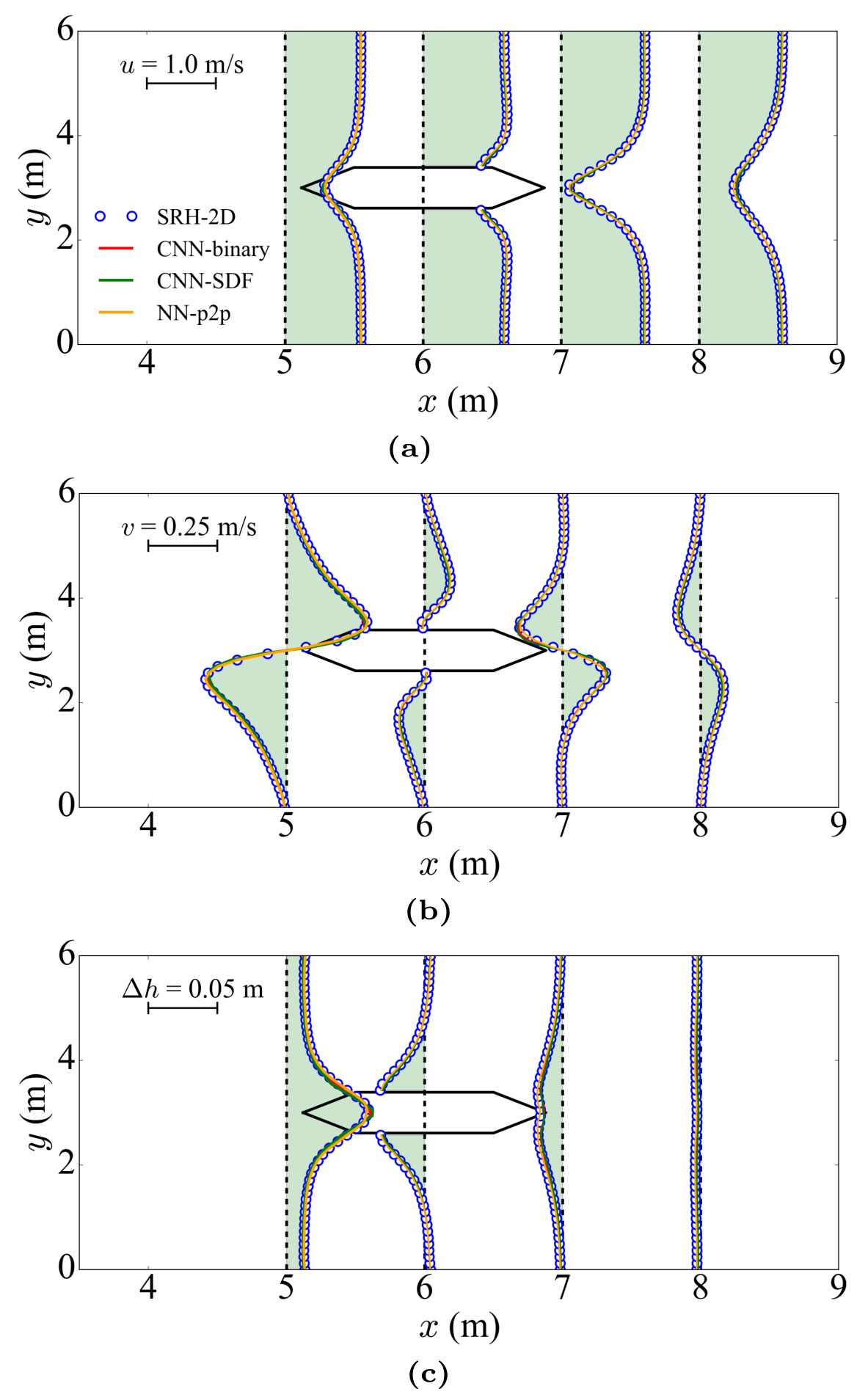}    
    \caption{Profiles of  flow variables at $x$ = 5, 6, 7, 8 m for the triangular-nosed pier case. (a) $u$, (b) $v$ , (c) $\Delta h$.}\label{Fig:tri_nosed_pier_profiles}
\end{figure}

As a first step to visually compare the performance of the three surrogate models, Fig.~\ref{Fig:tri_nosed_pier_prediction} shows contour plots of flow field from different models in comparison with the ground truth (SRH-2D result). The case is the same as the one shown in Fig.~\ref{Fig:tri_nosed_pier_simulation}. Qualitatively, the general features of the flow field are captured by all three surrogate models.  When flow approaches the pier in the channel, the flow passing around the two sides of the pier contracts, accelerates, and causes the decrease of water depth. As the flow expands again in the downstream region, the water level gradually recovers to the specified value at the downstream boundary. To further check the results, Fig.~\ref{Fig:tri_nosed_pier_profiles} plots the flow variable profiles at $x$ = 5, 6, 7, and 8 m from the SRH-2D simulation and the three surrogate models. Again, the difference among them are too small to be discerned. 
 
\begin{figure}[htp]
\hspace{-50pt}
    	\includegraphics[width=1.2\textwidth]{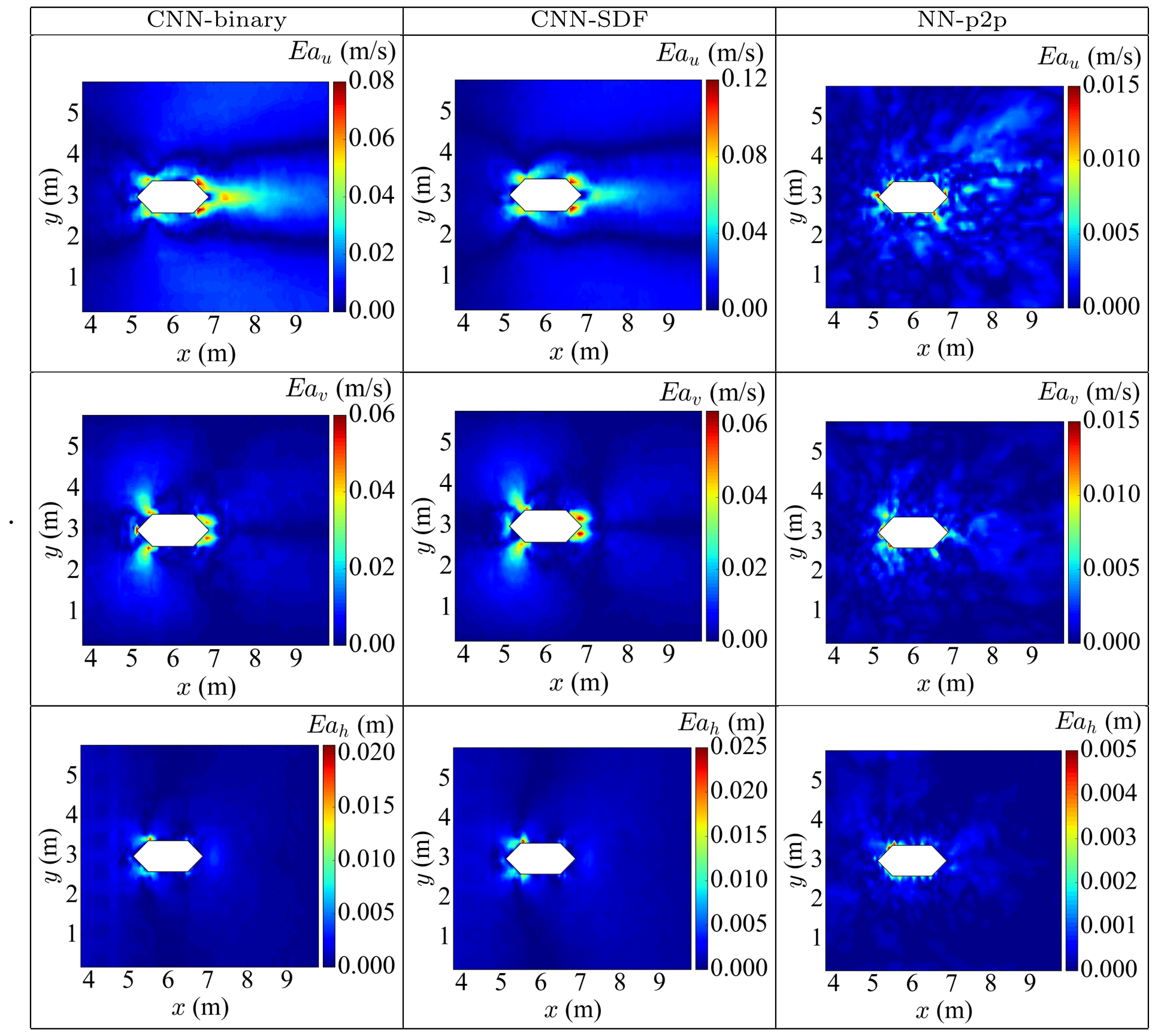}   
  \caption{Absolute errors in the results from the three surrogate models: CNN-binary, CNN-SDF, and NN-p2p.}\label{Fig:tri_nosed_pier_error}
\end{figure}

The errors can better show the performance of surrogate models. Figure~\ref{Fig:tri_nosed_pier_error} shows the absolute error contours of the predicted flow fields. For all surrogate models, most errors occur near the edge of the pier. The maximum absolute errors of ${u}$ and ${v}$ predicted by CNN-based models are around 0.12 m/s and 0.06 m/s, respectively, and the maximum absolute error of water depth is around 0.025 m. However, the maximum absolute errors of velocity components predicted by NN-p2p are less than 0.015 m/s and the maximum absolute error of water depth is less than 0.005 m. From the figures, it is clear that NN-p2p surrogate model has great advantages in capturing the flow in the vicinity of the pier where the flow field undergoes rapid transition due to the presence of the solid boundary. The reason is that in NN-p2p, the boundary information is in vector format (points) and exact. The neural network learns directly on this information, in contrast to the indirect information embedded in the raster images used by CNN-based methods. In addition, the contour plots show that the largest errors in the CNN-based methods are located near the corners of the pier. This is again due to the inability for raster images to accurately represent sharp corners.

To evaluate the overall performance of the three surrogate models for a wide range of cases, Table~\ref{Tab: average_relative_error} shows the summary of the average absolute ($Ea$) and relative ($Er$) errors in the training domain of all testing cases. It is clear that the new NN-p2p method greatly reduces the errors for all flow variables. The performances of CNN-binary and CNN-SDF methods are comparable, but significantly lower than that of NN-p2p. It is noted that the relative errors for $v$ and $\Delta h$ from all three methods are high ($>$ 5\%). However, this should not cause any alarm because the absolute errors are very minimal. The high relative error is because the magnitudes of $v$ and $\Delta h$ are small and thus the denominator in Eqn.~\ref{eqn:errors} is small. 

\begin{table}[htp]
\caption{The average absolute and relative errors in the training domain for all testing cases.}
\small
\centering 
\begin{adjustwidth}{-1cm}{}
\begin{tabular}{c| c| c| c| c| c| c|c|c} 
\hline
\multirow{2}{*}{Flow variable} &\multicolumn{2}{c|}{$ {u} $}&\multicolumn{2}{c|}{${v}$}&\multicolumn{2}{c|}{${\sqrt{u^2+v^2}}$}&\multicolumn{2}{c}{$ \Delta h$}\\ \cline{2-9}
{} & $Er$ (\%)  & $Ea$ (m/s)& $Er$ (\%) & $Ea$ (m/s) &$Er$ (\%) & $Ea$  (m/s) &$Er$ (\%) & $Ea$ (m)\\
\hline
CNN-binary &  3.44 &0.017 &11.5 &0.007 & 2.87 & 0.018& 13.79 & 0.0295 \\
\hline
CNN-SDF &  3.74 &0.019  &12.2 &0.008 & 3.07 & 0.019& 15.98 & 0.0298 \\
\hline
NN-p2p  &  0.77 &0.003  &7.2 &0.002& 0.56 & 0.003& 5.70 & 0.0004 \\
\hline
\end{tabular}
 \end{adjustwidth}
\label{Tab: average_relative_error}
\end{table}

\subsection{Spatial extrapolation from the surrogate models}
The surrogate models seem perform relatively well in the training domain. How they perform in the outside areas (see Fig.~\ref{Fig:mesh}) is another indicator for how much physics the models learn. For practical purpose, we are interested in how far the surrogate model's results can be extrapolated upstream and downstream. Note that the outside areas, where the SRH-2D simulation data were not seen by the neural networks, are the regions where $x$ is in the range of 0 m to 3 m and 10 m to 15 m.

\begin{figure}[htp]
\hspace{-100pt}
  	\includegraphics[width=1.5\textwidth]{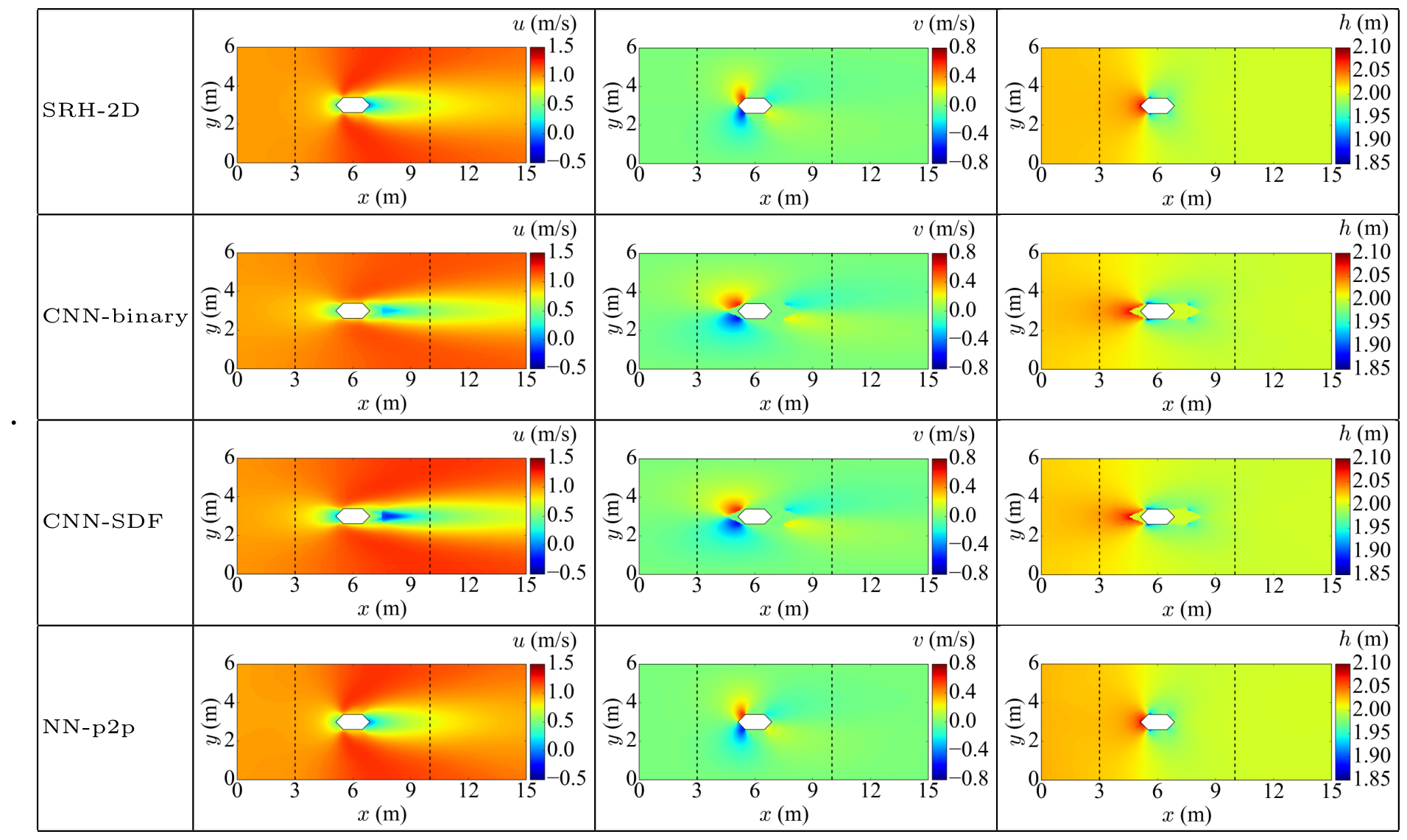} 
  \caption{Velocity and water depth contours around a triangular-nosed pier in the whole domain from SRH-2D simulation, CNN-binary, CNN-SDF and NN-p2p. The training domain is bounded by the two vertical dashed lines.}\label{Fig:tri_nosed_pier_prediction_all_domain}
\end{figure}

Figure~\ref{Fig:tri_nosed_pier_prediction_all_domain} shows the flow fields predicted by SRH-2D simulation, CNN-binary, CNN-SDF, and NN-p2p in the entire domain. CNN-based methods produce results with significant errors in the outside areas. This is in fact not surprising because in CNN-based methods the input image size (width and height) is fixed. To spatially extrapolate, the input images have to fit the enlarged domain by scaling, which distorts the pier boundary. From Fig.~\ref{Fig:tri_nosed_pier_prediction_all_domain}, CNN-based methods seem to only ``extend'' the flow data at the two vertical lines to upstream and downstream, not physically predict. This is why the wake zone predicted by the CNN-based methods is too elongated in comparison with the true solution. In conclusion, CNN-based methods are not suitable for spatial extrapolation because of their structural requirement.

On the other hand, the proposed NN-p2p method can produce reasonable predictions in regions not included in the training set. It is evident from Fig.~\ref{Fig:tri_nosed_pier_prediction_all_domain} that NN-p2p predicts the flow field in the outside areas well and it can properly capture the wake dynamics beyond the training domain. The advantage of NN-p2p is that it only requires the spatial coordinates $x$ and $y$ as input, not a fixed size image. Therefore it is flexible to be used for spatial extrapolation. The maximum absolute errors from NN-p2p are located near the outlet, which makes sense because the farther away from the training domain, the larger the extrapolation error. Nevertheless, the maximum absolute errors for $u$, $v$ and $h$ extrapolation are 0.04 m/s, 0.015 m/s and 0.005 m, respectively, which are very small considering their background values. It is also noted that despite the success shown here, extrapolation should always be used with caution and the extrapolated results should always be checked before used.


\subsection{Validations with physical laws}
In PBMs, the physical laws are directly and strictly enforced. However, in surrogate models these physical laws are embedded in the data and only indirectly enforced. Therefore, how well the surrogate models respect the physical laws is another important factor to consider. To demonstrate, we performed a simple check on the conservation of mass. Similar check on the conservation of momentum can be performed. At steady state, the total water flux across the boundary of a cell should be zero to satisfy mass conservation. The total water flux for a cell can be written as
\begin{equation}
\Delta m = \int_s h\mathbf{u} ds = \sum_{edges} h\mathbf{u} \cdot \mathbf{n} l_{edge}
\end{equation}
where $s$ is the boundary of a cell, $\mathbf{u}$ is the velocity vector $(u, v)$, $\mathbf{n}$ is the out normal vector of cell's edges, and $l_{edge}$ is edge's length.

\begin{figure}[H]
\centering
 \includegraphics[width=0.7\textwidth]{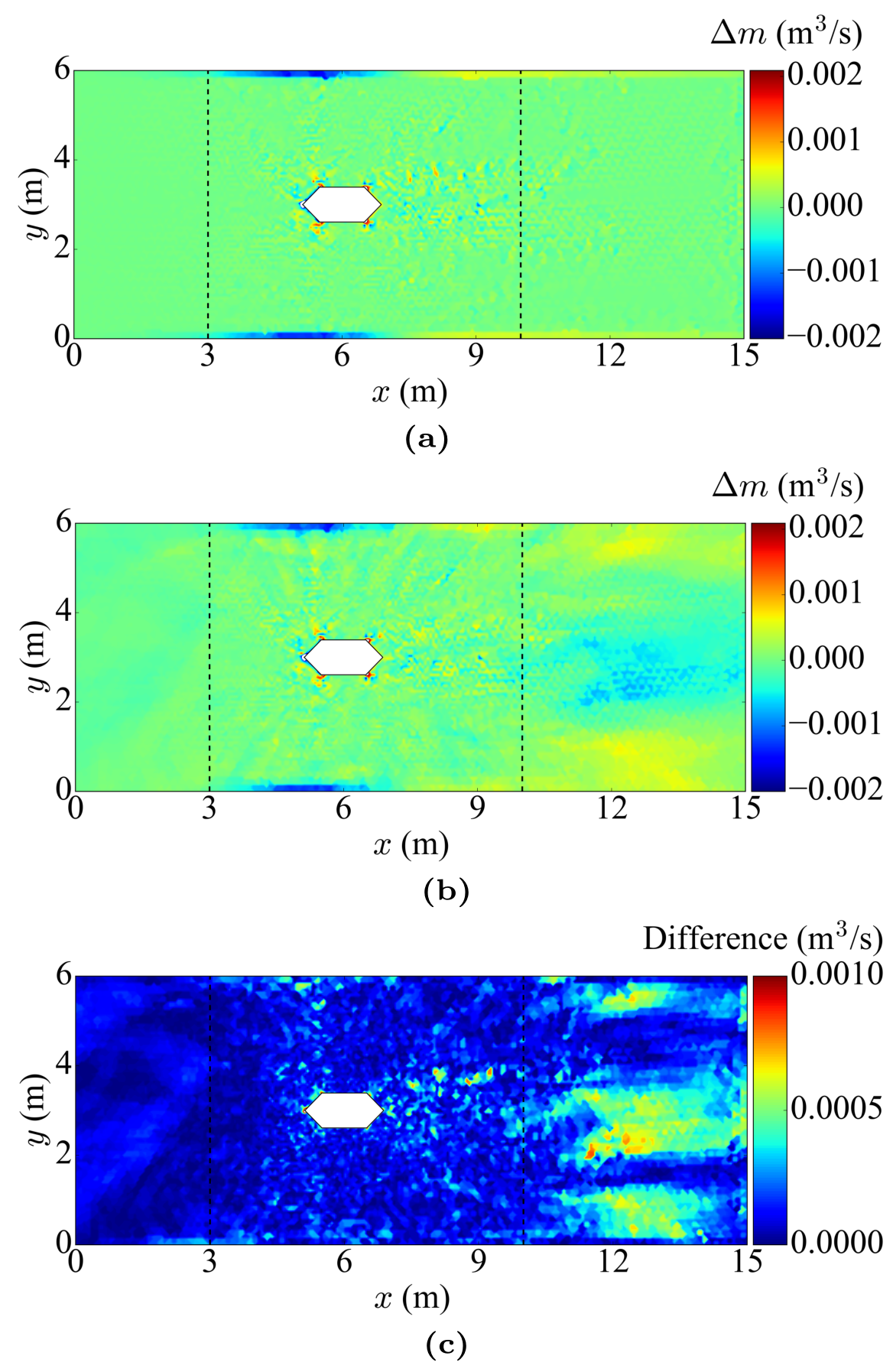}     
    \caption{Mass conservation errors in the entire domain: (a) SRH-2D, (b) NN-p2p, and (c) the difference in errors from SRH-2D and NN-p2p.}\label{Fig:mass_change}
\end{figure}

Figure~\ref{Fig:mass_change} shows the contours of mass conservation error $\Delta m$ in the entire domain from SRH-2D and NN-p2p. The conservation error in SRH-2D is caused by round-off errors and truncation errors that exist in any numerical models. The NN-p2p prediction satisfies mass conservation well and its error magnitude is comparable to that from SRH-2D. It is also observed that the mass conservation error from NN-p2p is larger in the outside areas than in the training domain. Again, this is due to spatial extrapolation. 

\subsection{Example applications of surrogate models}
\subsubsection{Drag coefficient}
\begin{figure}[htp]
\hspace{-2.5cm}
     \subfloat[][]{
    	\includegraphics[width=0.525\textwidth]{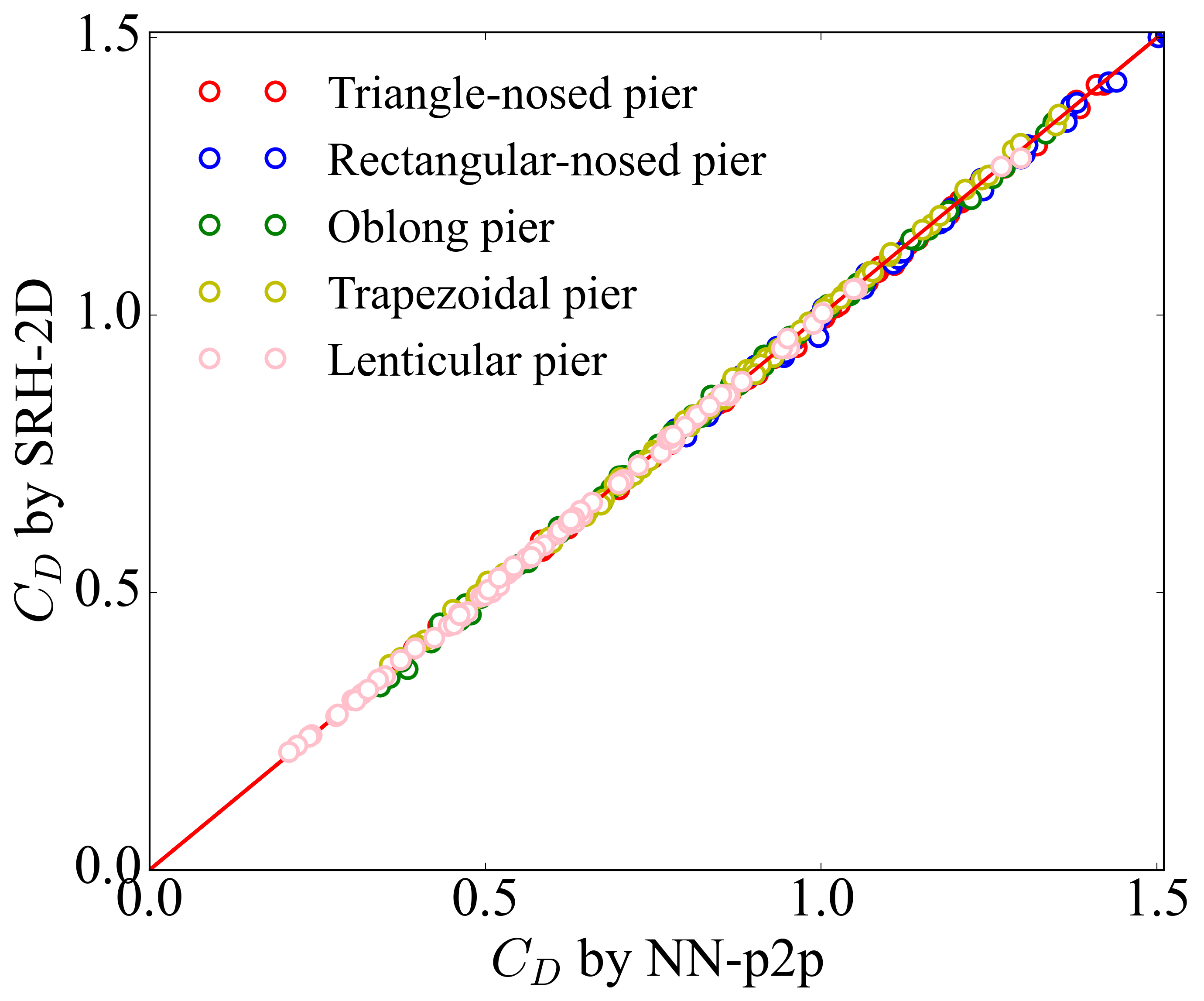}
     }   
     \subfloat[][]{
    	\includegraphics[width=0.9\textwidth]{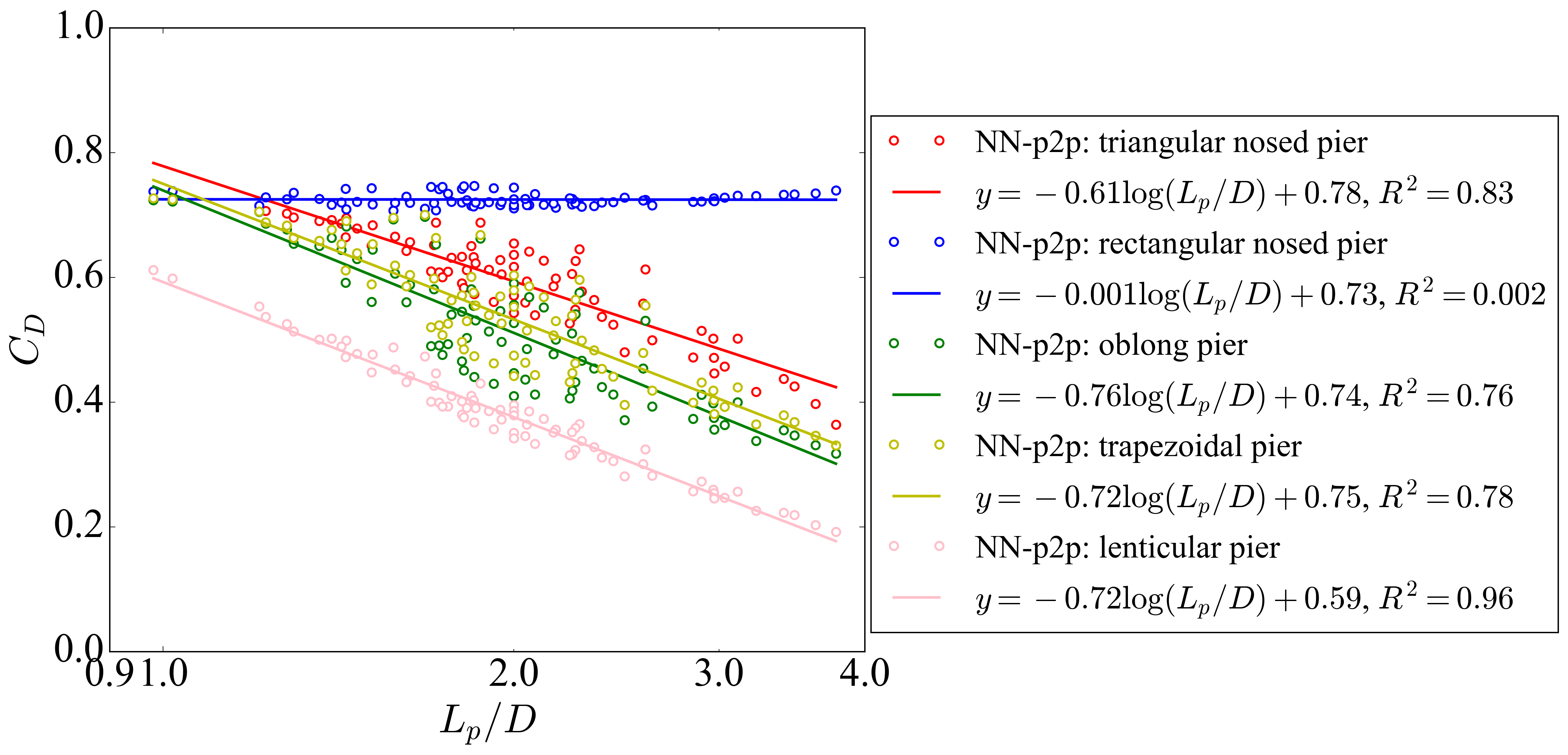}
     }        
    \caption{Drag coefficients for cases in the testing dataset: (a) comparison of $C_D$ by SRH-2D and NN-p2p, (b) $C_D$ scales linearly with $\log(L_p/D)$. }\label{Fig:CD_test}
\end{figure}

The drag coefficient, $C_D$, is an important parameter in the design of bridge piers against floods. The $C_D$ values for all cases in the testing dataset from SRH-2D and NN-p2p are compared in Fig.~\ref{Fig:CD_test} (a). An almost perfect match between the simulation and the NN-p2p model is obtained. This is in fact because the NN-p2p model is able to almost perfectly predict the water depth (pressure) at the grid points on the pier boundary. It is well known that the drag coefficient is a function of the pier shape and size. Using the $C_D$ values from NN-p2p, we found that $C_D$ decreases with the increase of $L_p/D$, the ratio of pier length $L_p$ to width $D$ except for the rectangular nosed pier, which has an almost constant value for all the cases simulated. We also found that there is a semi-linear relationship between $C_D$ and $\log(L_p/D)$. The linear regression results between $C_D$ and $\log(L_p/D)$ for all pier shapes are shown in Fig.~\ref{Fig:CD_test} (b). For most cases, relatively high $R^2$ values were obtained, indicating the robustness of the linear relationship.

\subsubsection{Unseen pier shapes}
\begin{figure}[htp]
\centering
     \subfloat{
    	\includegraphics[width=0.6\textwidth]{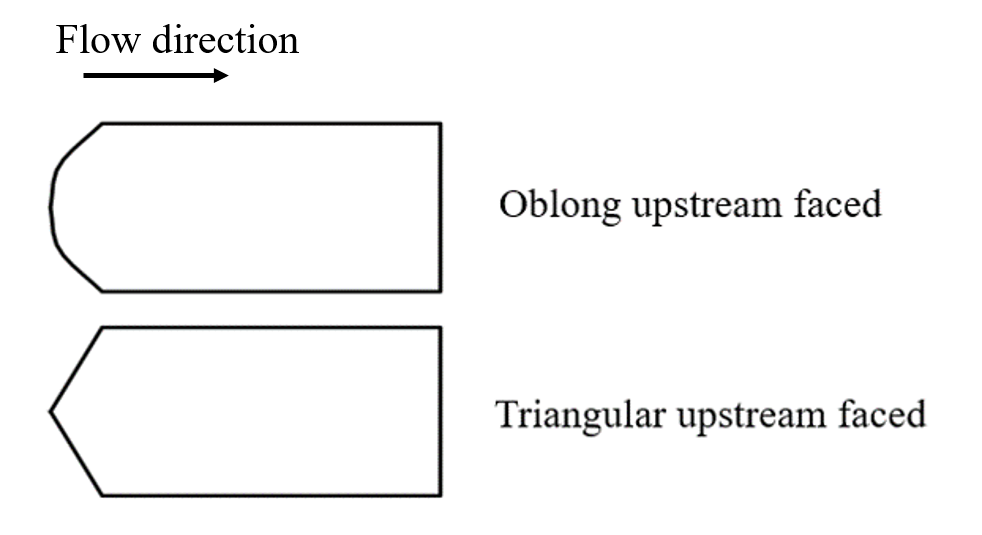}
     }   
    \caption{Two new pier shapes unseen by NN-p2p.}\label{Fig:new_piers}
\end{figure}

The true value of surrogate models is for them to predict flow around pier shapes unseen by the neural networks during training. To this end, two new pier shapes were tested: oblong upstream faced pier, and triangular upstream faced pier (Fig.~\ref{Fig:new_piers}). Note that these two shapes are different from those in Figure~\ref{Fig:Piers_boundaries} because they only have upstream noses. They have the same pier length and width as the testing dataset.

\begin{figure}[htp]
\hspace{-100pt}
 \includegraphics[width=1.5\textwidth]{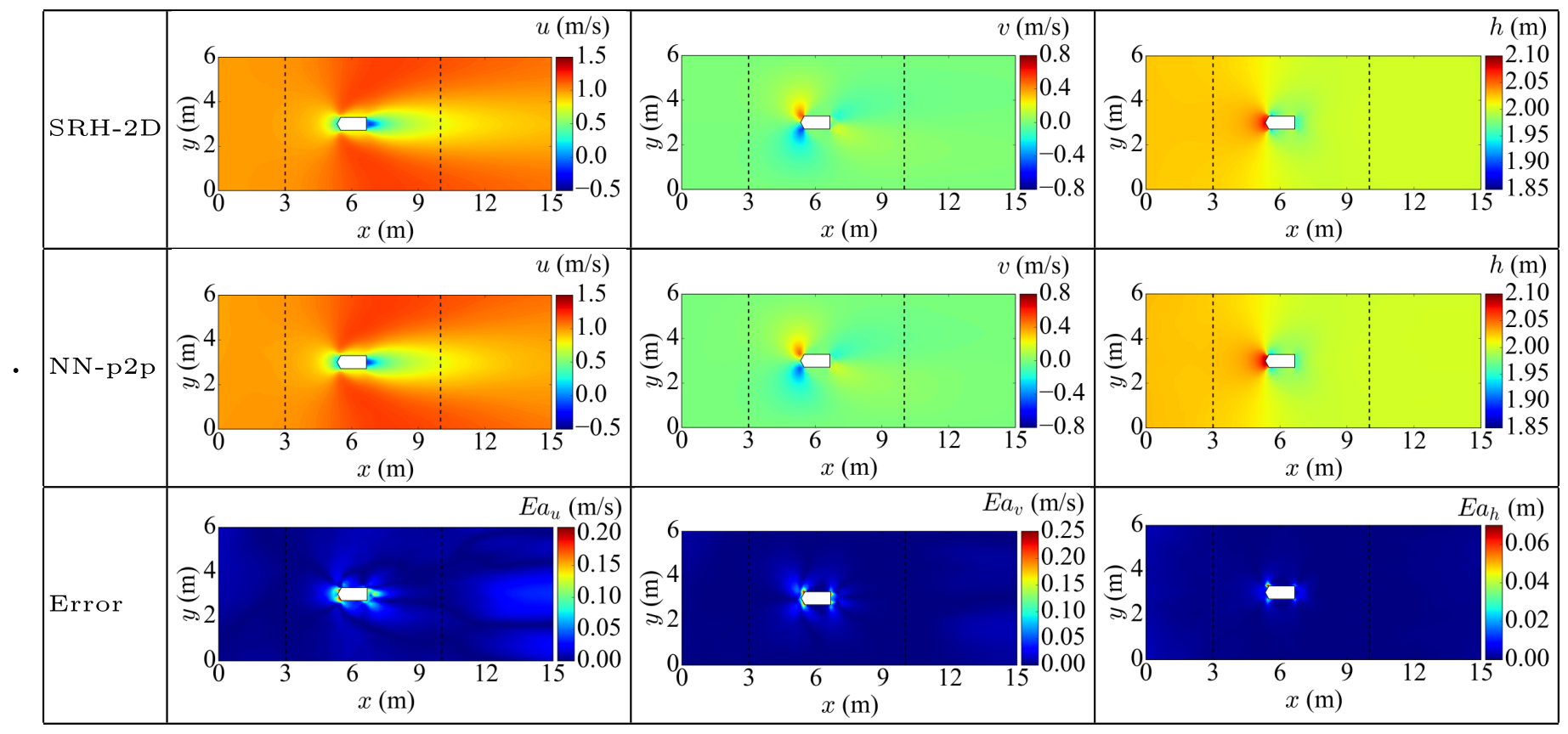}     
  \caption{Flow and water depth contours around a triangular upstream faced pier from SRH2D simulation and NN in the entire simulation domain.}\label{Fig:tri_rect_pier_prediction_all_domain}
\end{figure}

The flow field predicted by NN-p2p was compared against the ground truth from SRH-2D. As an example, Fig.~\ref{Fig:tri_rect_pier_prediction_all_domain} shows the comparison of flow fields around the triangular upstream faced pier in the entire domain. The errors by NN-p2p are also shown. In general, NN-p2p predicts well the flow field around unseen piers, even in the far upstream and downstream regions where no training data were provided. The maximum errors for the three flow variables, located around upstream corners, are very small comparing to their respective background values. 


%
%

To further quantify the errors in the predictions for unseen piers, Table~\ref{Tab: average_relative_error_new_shapes} shows the calculated errors, which are all in reasonably small ranges. It is found that the streamwise velocity, ${u}$, is better learned judging by the small relative error, perhaps due to the fact it is the dominant flow variable in the problem. The relative errors of ${v}$ and $\Delta h$ are much larger than that of ${u}$, the absolute errors of ${v}$ and $\Delta h$ are much smaller than ${u}$ in magnitude. It is perhaps more revealing in using the error for the velocity magnitude $\sqrt{u^2+v^2}$, which has a very small error.

\begin{table}[h]
\caption{The average absolute and relative errors for the two new pier shapes predicted by NN-p2p.}
\centering 
\small
 \begin{adjustwidth}{-2cm}{}
\begin{tabular}{c| c| c| c| c| c| c|c|c} 
\hline
\multirow{2}{*}{Flow variables} &\multicolumn{2}{c|}{${u} $}&\multicolumn{2}{c|}{${v}$}&\multicolumn{2}{c|}{$\sqrt{u^2+v^2}$}&\multicolumn{2}{c}{$ \Delta h$}\\ \cline{2-9}
{} & $Er$ (\%)  & $Ea$ (m/s)&$Er$ (\%) & $Ea$ (m/s) &$Er$ (\%) & $Ea$  (m/s) &$Er$ (\%) & $Ea$ (m)\\
\hline
Oblong upstream faced &  7.6\% &0.043 &35.5\% &0.027  &7.3\% &0.040 & 23.9\% & 0.005  \\
\hline
Triangular upstream faced &  4.9\% &0.028 &36.6\% &0.018  &4.9\% &0.027  & 26.0\% & 0.003  \\
\hline
\end{tabular}
\label{Tab: average_relative_error_new_shapes}
 \end{adjustwidth}
\end{table}

\begin{figure}[htp]
\hspace{-2.5cm}
     \subfloat[][]{
    	\includegraphics[width=0.525\textwidth]{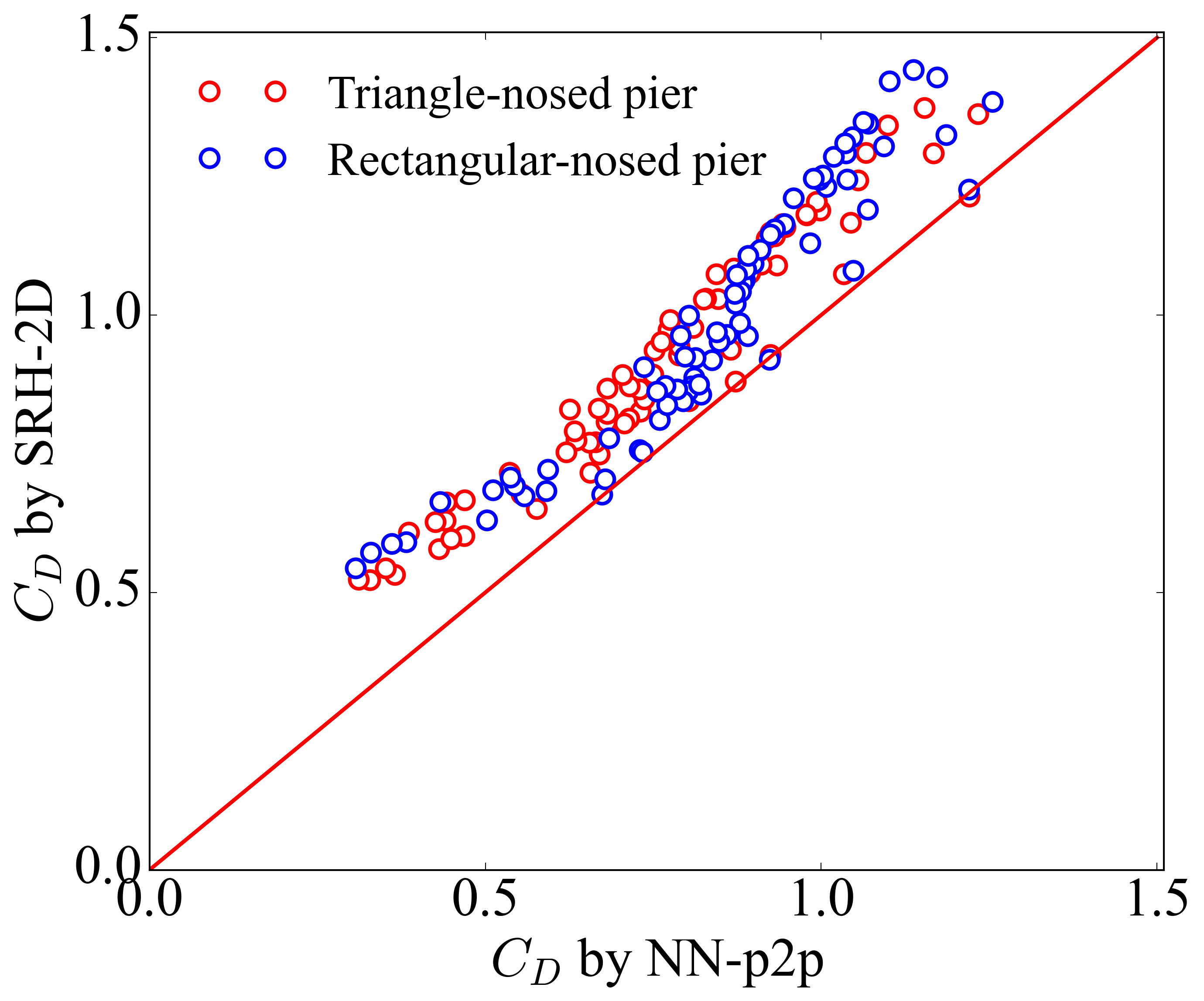}
     }        
     \subfloat[][]{
    	\includegraphics[width=0.9\textwidth]{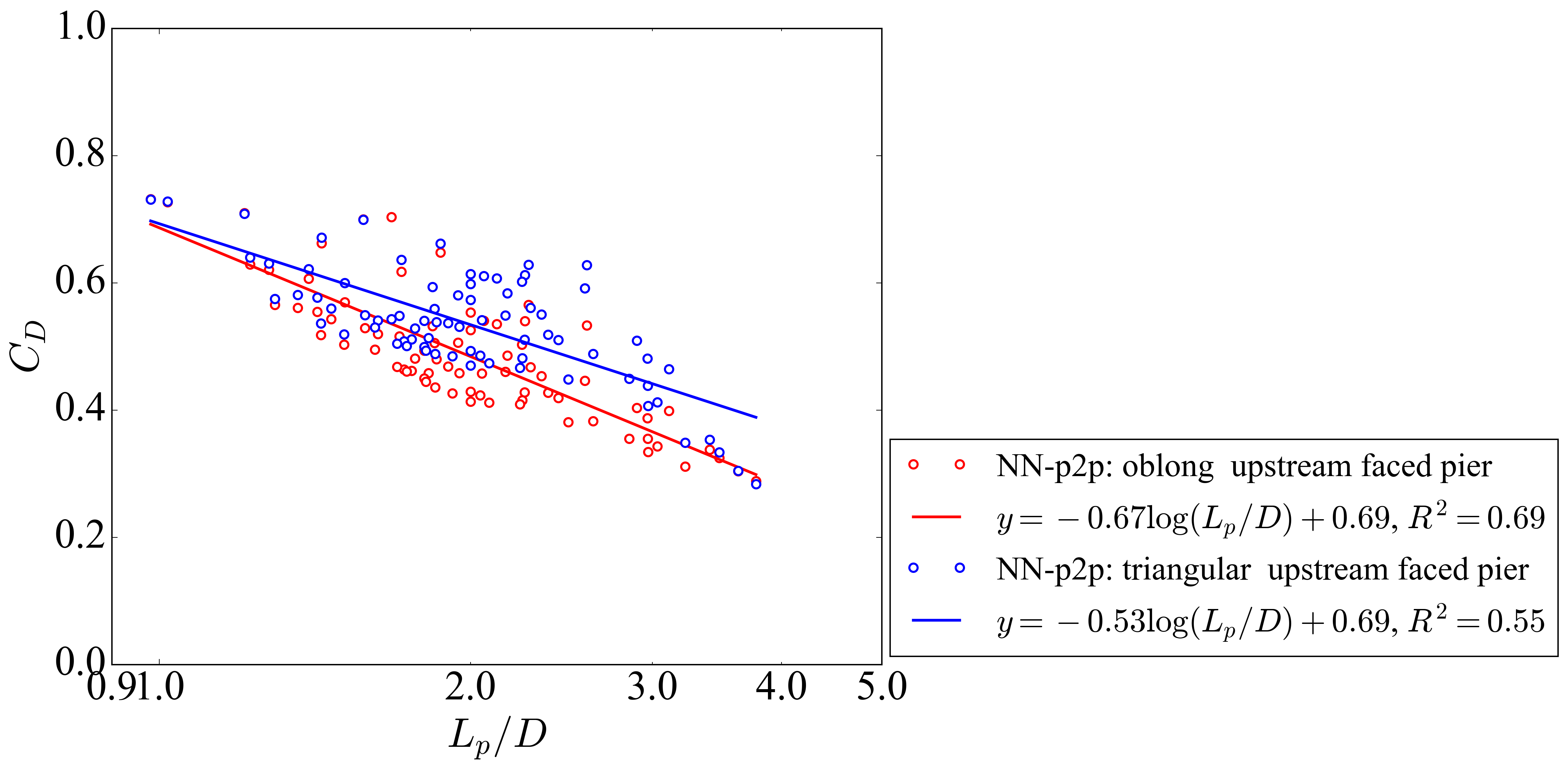}
     }       
    \caption{Drag coefficients for piers with new shapes: (a) comparison of $C_D$ by SRH-2D and NN-p2p, (b) $C_d$ scales linearly with $\log(L_p/D)$.}\label{Fig:CD_new_shape}
\end{figure}

As in previous section, the drag coefficients for the new piers were also examined using the proposed surrogate model. Figure~\ref{Fig:CD_new_shape} (a) is the comparison of $C_D$ values predicted by SRH-2D and NN-p2p. The performance of the surrogate model slightly deteriorates in comparison with that for piers seen by the neural network. However, the difference in the $C_D$ values by SRH-2D and NN-p2p is still reasonably small. Furthermore, based on the results from NN-p2p, the linear relationship between $C_D$ and $\log(L_p/D)$ still holds as shown in Fig.~\ref{Fig:CD_new_shape} (b).

\subsection{Computational cost}
Reduction of computational cost is the main motivation of developing surrogate models. To appraise the time cost of all models used in this work, Table.~\ref{Tab: computational_time} shows the computing time by SRH-2D and the three surrogate models. It is clear that once trained, the surrogate models are much faster than SRH-2D by almost three orders of magnitude. Comparing the three surrogate models, it is found that the training of CNN-based models and their use for prediction are faster than those for NN-p2p. Nonetheless, CNN-based models are less accurate than NN-p2p and have limitations in extrapolation as discussed in this work.

\begin{table}[htp]
\caption{Computational time for tasks and models.}
\centering 
\begin{tabular}{l| c| c} 
\hline
\multirow{2}{*}{Tasks and Models} & CPU time & GPU time\\ \cline{2-3}
{} & Intel Core 3.40GHz & NVIDIA K80 \\
\hline
Averaged simulation time per case by SRH-2D & 600 s (0.17 h) & - \\ \hline
Training time of CNN-binary & - & 0.46 h \\ \hline
Training time of CNN-SDF & - & 0.83 h \\ \hline
Training time of NN-p2p & - & 1.46 h \\ \hline
Averaged prediction time per case by CNN-binary & 0.14 s & - \\ \hline
Averaged prediction time per case by CNN-SDF & 0.16 s & - \\ \hline
Averaged prediction time per case by NN-p2p & 1.33 s & - \\
\hline
\end{tabular}
\label{Tab: computational_time}
\end{table}

\section{Conclusion}
This work developed a surrogate model for the solution of shallow water equations with deep learning. The neural network was trained with point data and boundary features to precisely define the domain geometry. The new surrogate model makes point-to-point predictions for flow field, thus named NN-p2p. For comparison, two CNN-based methods, CNN-binary and CNN-SDF, were also implemented, where the input and output are raster images. All methods were used to simulate flow around bridge piers in a channel. The training data were generated from the 2D hydraulics model SRH-2D. Even though the CNN-based methods have reported high accuracy predictions on structured/regular meshes in the literature \citep{guo2016convolutional,zhu2018bayesian,bhatnagar2019prediction,ribeiro2020deepcfd}, NN-p2p still has advantages in the prediction on unstructured/irregular meshes. It produces more accurate flow field very close to the pier because of its capability to precisely define the pier geometry using vectorized information. In contrast, the accuracy of CNN-based methods is greatly limited by image structure and resolution. 

Several important questions regarding the feasibility, applicability and accuracy of the surrogate models are answered. For the question of whether they can predict the flow around seen piers but unseen dimensions, the surrogate models all performed relatively well and the difference in the predicted flow fields can not be visually discerned. Quantitatively, the calculated errors show the new NN-p2p greatly reduces the errors for all flow variables and performs better than CNN-based methods. The proposed NN-p2p model also works well to predict flow around piers with unseen shapes. Flow around two new pier shapes was predicted and the error against SRH-2D was small. 

For the question of whether they can accurately predict the flow outside the zone of training data, the surrogate models were used to make spatial extrapolations to obtain flow fields in regions outside of the training domain. Limited by the structural design of CNN-based methods, it is very challenging to expand their prediction to the domain not covered by the training images. In our case, the extrapolated flow fields from CNN-based methods completely miss the flow physics such as wake behind object. On the other hand, NN-p2p is well suited to perform spatial extrapolation and it can reasonably predict wake dynamics.

The prediction of NN-p2p surrogate model respects physical laws in the training data. For example, the mass conservation for each cell in the mesh was calculated and the error is very small and comparable with the numerical error in SRH-2D. In areas of spatial extrapolation, the mass conservation error slightly increases. 

With the confidence gained in the performance of the NN-p2p method, some applications were demonstrated. For example, the flow results from NN-p2p were used to calculate the drag coefficient $C_D$ and a semi-linear relationship was found between $C_D$ and $\log(L_p/D)$, the log transform of a pier's length to width ratio.

In summary, the proposed NN-p2p surrogate model can efficiently and accurately solve the shallow flow around objects in open channels. The solution has comparable accuracy as physics-based models such as SRH-2D. Furthermore, the NN-p2p surrogate model is very flexible and the computational cost is negligibly low, which makes it ideal for real time prediction and parameter inversion. 

 \bibliographystyle{elsarticle-harv} 
 \bibliography{references}





\end{document}